\documentclass[10pt, aps,prc,twocolumn,superscriptaddress,preprintnumbers,
amsmath, 
floatfix,
longbibliography,
nofootinbib
]{revtex4-1}
\usepackage[T1]{fontenc}
\usepackage[utf8x]{inputenc} 
\usepackage{adjustbox}          
\usepackage[caption=false]{subfig}
\usepackage{url}
\usepackage{color}
\usepackage{float}
\usepackage[pdftex,colorlinks=true, linkcolor = blue, citecolor=blue,urlcolor=blue, bookmarksnumbered=true, bookmarksopen=true]{hyperref}
\usepackage{amsfonts}
\usepackage{wrapfig,bm,bbm}
\newcommand{\beq}{\begin{equation}}
\newcommand{\eeq}{\end{equation}}
\newcommand{\bea}{\begin{eqnarray}}
\newcommand{\eea}{\end{eqnarray}}

\begin{document}
\title {\bf  Framework for Polarized Superfluid Fermion Systems }
  
\author{Aurel Bulgac}
\email{bulgac@uw.edu}
\affiliation{Department of Physics, University of Washington, Seattle, Washington 98195--1560, USA}
  
\date{\today}

\begin{abstract}
I discuss the advantages and disadvantages of several procedures, some known and some new, for 
constructing stationary states within the mean field approximation for a system with pairing correlations 
and unequal numbers spin-up and spin-down  fermions, using the two chemical potentials framework. 
One procedure in particular appears to have significant physics advantages over previously
suggested in the  literature computational frameworks. Moreover, this framework is applicable 
to study strongly polarized superfluid fermion systems with arbitrarily 
large polarizations and with arbitrary total particle numbers. These methods are equally applicable to 
normal systems.

\end{abstract}

\preprint{NT@UW-20-25}

\maketitle

\providecommand{\selectlanguage}[1]{}
\renewcommand{\selectlanguage}[1]{}

\section{Introduction} 

It is well known that by removing or adding a nucleon from a magic nucleus the
self-consistent mean field solution breaks the rotational symmetry and time-odd
contributions also appear in the mean field. By adding or removing more nucleons
typically nuclei become more deformed in the mean field description and their excited 
spectra evolve from vibrational to rotational spectra~\cite{Bohr:1969,Ring:2004}.  
One can ask the question: 
``How a large spin polarization of nuclei, neutron or nuclear matter would modify 
the properties of their ground states and spectra and their response to various external 
perturbations?" The interest in spin polarized nuclei, neutron, and nuclear 
mater is extensive~\cite{Bauer:2020,Polls:2007,Urban:2020,Riz:2018,
Stein:2018,Stein:2016,Forbes:2014,Tews:2020,Torres-Patino:2019,
Behera:2016,Dexheimer:2017,Harutyunyan:2016,Rabhi:2015, Stein:2016a,
Sammarruca:2015,Behera:2015,Vidana:2016,Lacroix:2015,Kruger:2015,
Aguirre:2014,Bordbar:2013,Dong:2013,Isayev:2011,Isayev:2012,Isayev:2006}.

The odd or the odd-odd superfluid nuclei are the simplest examples
of spin polarized systems. The definition 
of a spin polarized system depends on various circumstances. Polarization, in 
particular the magnetization of condensed matter systems, can be due either to a
spontaneous symmetry breaking or it can be induced by 
an external field. The magnitude of the polarization, or loosely speaking 
of the difference between the particle numbers with spins up and down 
$N_\uparrow-N_\downarrow$ is not as a rule an integer. Moreover, 
$N_\uparrow-N_\downarrow\neq 0$
can occur in both normal and superfluid systems. Polarized systems can be also 
artificially ``manufactured'' and this is routinely done in cold atom physics.  

In a time-dependent framework, a major
difficulty in studying the dynamics of a superfluid fermionic system
in general or in particular of a nuclear system with an imbalance of
spin-up and spin-down fermions ($N_\uparrow \neq N_\downarrow$), is
the construction the initial stationary state needed in time-dependent
treatments.  Using the simple blocking approximation for the initial
state could be problematic in the case of time-dependent
phenomena. Within the blocking approximation the odd fermion is often
described by a single-particle wave function and not by a Bogoliubov
quasiparticle wave function, and thus the orthogonality between the
two types of fermion wave functions cannot be enforced during the time
evolution, except within the BCS approximation. Using the equal
filling approximation, when the extra odd fermion occupation
probability is spread uniformly among several levels leads to
unphysical polarization properties of the system, which can have major
unphysical consequences when describing the nuclear fission of odd or
odd-odd nuclei.
 
Of particular interest is the microscopic description and the
understanding of the nuclear fission dynamics, particularly of the
stage where the fissioning nucleus emerges from below the potential
barrier near the saddle, to the scission configuration. This part 
of the nuclear evolution  is a
highly non-equilibrium process, where pairing correlations play
the role of a very efficient
``lubricant''~\cite{Bertsch:1997,Bulgac:2019b,Bulgac:2020}.
Experiments show that the fission of odd or odd-odd nuclei is hindered
significantly when compared to the fission of the neighboring
even-even nuclei~\cite{Vandenbosch:1973,Wagemans:1991}. The unpaired
fermion(s) can be found in a state with a relatively high total
orbital angular momentum projection on the reaction axis and in that
case the presence of an odd fermion can have a strong hindering
effect, as the pairing correlations, particularly within the
traditional theoretical nuclear approaches used so far, are
ineffective in the case of the extra nucleon.  This is very easy to
understand. In the case of an odd or odd-odd heavy nucleus the odd
particle(s) can have a very large orbital angular momentum along the
symmetry axis $l_z\approx \pm p_FR\approx 9 \hbar$, where $p_F$ is the
Fermi momentum and $R\approx 1.1A^{1/3}$ is the waist radius of the
fissioning nucleus.  Such single-particle states are populated for
example when studying the induced fission of $^{239}$U obtained in the
one-neutron transfer reaction, such as the recently published
experimental study
$^9$Be($^{238}$U,$^{239}$U)$^8$Be~\cite{Ramos:2020}.  In the
fission fragments the maximum angular momentum is expected however to be
$l_z(H)\approx \pm 7.5\hbar$ and $l_z(L)\approx \pm 6.7\hbar $ for the
heavy (H) and light (L) fragments respectively, thus smaller by
roughly $15\pm5$\% than in the mother nucleus. Since nuclei in the
majority of theoretical models used so far are predominantly axially
symmetric along the fissioning
direction~\cite{Bender:2004,Ryssens:2015,Schunck:2016}, when they
fission the value of $l_z$ of the odd nucleon(s) should be conserved,
as the pairing correlations are ineffective in lowering 
the angular momentum of the extra nucleon or
nucleons. This is unlike the case of an even-even nucleus where pairing is
very efficient in performing transitions of the type
$(j_z,-j_z)\rightarrow (j'_z,-j'_z)$ and can thus lower the orbital
angular momentum of nucleon
pairs~\cite{Bertsch:1997,Bulgac:2019b,Bulgac:2020}.  Clearly the
maximum initial $l_z\approx \pm 1.1A^{1/3}p_F$ and the maximum final
values of the odd nucleon(s) in the heavy and light fragments
$l_z(H,L)\approx \pm 1.1(A_{H,L})^{1/3}p_F$ differ vastly, a tension
which is usually at the root of the argument why fission of odd and
odd-odd nuclei is hindered when compared with the fission of even-even
nuclei. The breaking of the axial symmetry obviously can help, but
that woulds require some deformation energy. Also collisions, which
are neglected in mean field treatments, can also help to transfer
several units of angular momentum, but they are Pauli suppressed at
low intrinsic excitation energies. Both of these processes are
expected to be less efficient that the pairing correlations, which are
can be enhanced due to the presence of a pairing
condensate~\cite{Bulgac:2019b,Bulgac:2020}.
 
One has to remember that the $s$-wave pairing type of matrix elements,
which are by design included within a time-dependent Density
Functional Theory (DFT) for superfluid
systems~\cite{Bulgac:2013a,Bulgac:2019,Bulgac:2019c}, account for a
significant portion of the collision integral at low excitation
energies, even in the absence of a true pairing condensate. The main
difference between balanced/unpolarized ($N_\uparrow =N_\downarrow$)
and imbalanced/polarized ($N_\uparrow \neq N_\downarrow$) superfluids
is that in the first case the Cooper pairs have zero momentum and the
Coopers pairs have a finite momentum in the other case
respectively~\cite{Bulgac:2008,Bulgac:2011a}, a fact known
theoretically for a long time in condensed matter
systems~\cite{Larkin:1964,Fulde:1964} 
and in cold atom physics as well.  The presence of Cooper pairs
with finite momentum will be instrumental in lowering the orbital
angular momentum of the extra nucleons in case of nuclear fission.
  
One theoretical method used to describe (static) pairing correlations
in a system with an odd number of fermions requires the use of two
different chemical potentials for the approximately time-reversed
single-particle states, for spin-up and spin-down respectively, and
was discussed in
Refs~\cite{Bulgac:2007,aslda,Sensarma:2007,Bulgac:2008}.  The presence
of an odd fermion can lead to a relatively weak time-reversal mean
field symmetry breaking and to different mean fields for the partners
of the Cooper pair, induced by the polarization effects due to the odd
fermion. The two chemical potential approach has been used in
Refs.~\cite{Bulgac:2007,Sensarma:2007,Bulgac:2008,Bulgac:2011a,
Bertsch:2009,Wlazlowski:2018,Magierski:2019,Tuzemen:2019}.
\footnote{See also the actual MATLAB code used in those
  calculations~\cite{aslda}, using the DVR method described in
  Ref.~\cite{Bulgac:2013}, and where the simulated annealing method
  was used along with the Broyden method~\cite{Baran:2008} for the
  iterative process.}

The general approach used to describe odd fermion systems used in nuclear
physics~\cite{Dobaczewski:1997,Dobaczewski:2000a,Ring:2004,Schunck:2017,Bertsch:2009a}
has the major disadvantage that one needs to know {\it a priori} the
quantum numbers of the odd fermion.  \textcite{Robledo:2011} however
have shown that a gradient technique approach~\cite{Ring:2004} is
apparently free of this difficulty.  Sometimes the implementation of
the general approach is construed as gain, as one can determine at
once a slew of low lying excited states of the odd or of the odd-odd
nucleus, even though the computational price is high.  While adding a
single extra fermion to an even fermion system can appear as a small
perturbation $~{\cal O}(1/A)$, since the low energy spectrum of odd
and odd-odd nuclei is relatively dense, a small perturbation can in
principle lead to significant changes of the nuclear mean field.
 
The question I raise here is: ``Can one devise a more transparent and
computationally faster framework to describe pairing correlations in
odd and odd-odd nuclei in particular and for arbitrary spin-polarizations 
as well?'' Since in unpolarized fermion
systems, or in even-even nuclei one treats the spin-up and spin-down
fermions identically, naturally there is a need for only one
chemical potential for either neutrons or protons. It seems then that since in polarized or odd
fermion systems, or in odd and odd-odd nuclei, spin-ups and spin-down
clearly experience different mean fields, the introduction of two
chemical potentials does not appear to need a justification and it
seems like the most natural approach. In the end the framework I
describe can handle arbitrary spin-polarizations, not limited to the
$|N_\uparrow-N_\downarrow|=1$ case only, which can be of interest in a
number of situations.
  
In Section~\ref{sec:general} I briefly review the Bogoliubov transformation
and the definition of various densities for even and odd fermion
numbers.  In Section~\ref{sec:nuclei} I review a previously suggested
framework for odd fermion systems with axial and parity
symmetry~\cite{Bertsch:2009}, I also introduce a couple of generalizations
applicable when an octupole deformation is present, and I discuss
their advantages and disadvantages. In Section~\ref{sec:cold_atoms} I
review the framework designed for cold atom systems, in which case the
spin-orbit interaction is absent. In Section~\ref{sec:best} I
introduce the optimal two chemical potential framework, which appears
to be free of any of the disadvantages of those previously suggested
in the literature frameworks for nuclear systems.  In the
appendix~\ref{sec:Pauli} I discuss various representations of the
Pauli spin matrices and in the appendix~\ref{sec:numeric} I discuss
several aspects concerning the numerical implementation.
 
 \section{Formulation of Bogoliubov transformations for even and odd fermion numbers}\label{sec:general}
 
First I review the case of a system with an even number of fermions,
such as even-even nuclei.  The creation and annihilation quasiparticle
operators are represented as~\cite{Ring:2004}
\begin{align}
&\alpha_k^\dagger  = 
\int d\xi\left [ u_k(\xi) \psi^\dagger (\xi) + v_k(\xi) \psi (\xi)\right ], \label{eq:a0}\\
&\alpha_k= 
\int d\xi\left [ v_k^*(\xi) \psi^\dagger (\xi ) + u_k^*(\xi) \psi (\xi)\right ], \label{eq:b0}
\end{align}
and the reverse relations are
\begin{align} 
&\psi^\dagger (\xi) = \sum_k \left [ u^*_k(\xi)  \alpha^\dagger _k  
                                            + v_k(\xi)\alpha_k \right ], \label{eq:p1}\\
&\psi(\xi) =                \sum_k \left [ v^*_k(\xi)\alpha^\dagger_k
                                            + u_k(\xi)\alpha_k \right ], \label{eq:p2}
\end{align}
where $\psi^\dagger (\xi )$ and $ \psi (\xi)$ are the field operators
for the creation and annihilation of a particle with coordinates
$\xi=({\bm r},\sigma)$.  The normal number (Hermitian $n=n^\dagger$ )
and anomalous (skew symmetric $\kappa =-\kappa^T$) densities are
\begin{align}
&n(\xi,\xi') = \langle \Phi|\psi^\dagger(\xi')\psi(\xi)|\Phi\rangle \nonumber\\
&= \sum_k v_k^*(\xi) v_k(\xi')=\sum_{l=n,\bar{n}} v_l^2  \phi_l^*(\xi) \phi_l(\xi'),\label{eq:number} \\
&\kappa(\xi,\xi') = \langle \Phi|\psi(\xi')\psi(\xi)|\Phi\rangle \nonumber\\
& =\sum_k v_k^*(\xi)u_k(\xi') = \sum_{l=n,\bar{n}} u_lv_l \phi_l^*(\xi) \phi_{\bar{l}}^*(\xi'), \label{eq:anomal} \\
&  \int d\xi \phi_k^*(\xi) \phi_l(\xi)=\delta_{kl},
\end{align}
with $u_l^2+v_l^2=1$, $0\le u_l=u_{\bar{l}}\le 1$, $0\le
v_l=-v_{\bar{l}}\le 1$, and $n$ and $\bar{n}$ label the time-reversed
states $\phi_l(\xi), \phi_{\bar{l}}(\xi) $ in the canonical
representation~\cite{Bloch:1962,Ring:2004}, and where
\begin{align} & \alpha_k|\Phi\rangle =0, \quad \langle
\Phi|\alpha_k^\dagger = 0, \quad \langle
\Phi|\alpha_k\alpha_l^\dagger|\Phi\rangle =\delta_{kl}.
\end{align} There is in general no rule on how to separate the
quasiparticle operators $\alpha^\dagger_k,\alpha_k$ into creation and
annihilation ones, and one can rename/interchange any number of them,
and declare a number of creation operators annihilation operators and
{\it vice versa}. This is unlike the field operators
$\psi^\dagger(\xi),\psi(\xi)$, which are defined with respect to the
true vacuum, $\psi(\xi)|0\rangle \equiv 0$. Only by requiring that the
quasiparticle vacuum $|\Phi\rangle$ corresponds to the lowest, or
often to a local minimum, of the total energy of an average even
number of fermions, one can clearly distinguish between creation and
annihilation quasiparticle operators.

In the case of an odd number of fermions the ground state is defined
as $\alpha^\dagger_\mu|\Phi\rangle$~\cite{Ring:2004}, where $\mu$ is
an appropriately chosen {\it a priori} quasiparticle state, and thus
\begin{align}
\alpha_l\alpha_\mu^\dagger|\Phi\rangle = \delta_{l\mu}, \quad  \langle \Phi|\alpha_\mu \alpha_k^\dagger = \delta_{k\mu}.
\end{align} 
Since by definition $|\Phi\rangle$ corresponds to an
average even number of fermions $N$, the state
$\alpha^\dagger_k|\Phi\rangle$ should describe an odd number of
fermions $N\pm1$, and their corresponding particle parity is given by
$(-1)^N=+1$ or $(-1)^{N\pm 1}=-1$ respectively.  Note however, than
the state $\alpha^\dagger_k|\Phi\rangle$, where $|\Phi\rangle$ is defined in Eq.~\eqref{eq:phi},
does not automatically has an integer average odd number of fermions,
as the chemical potential, and therefore the quasiparticle
wavefunctions should be correspondingly adjusted. As
\begin{align}
|\Phi\rangle \propto \prod_k \alpha_k |0\rangle, \label{eq:phi}
\end{align}
(assuming that $\alpha_k|0\rangle \ne 0$ for any $k$, otherwise see
\textcite{Ring:2004}) \footnote{In the case of a finite dimensional
Hilbert space it could be problematic to establish if mathematically
$\alpha_k|\Phi\rangle\ne 0$, since when $\int d\xi |v_k(\xi)|^2$ is
smaller then the machine precision the corresponding annihilation
operators do not anti-commute anymore and the ordering of the terms in
product Eq.~\eqref{eq:phi} can lead to results differing by more than
just by a phase.}  the ground state of an odd fermion system is
therefore defined as
\begin{align}
\alpha_l^\dagger|\Phi\rangle \propto \prod_{k\ne l} \alpha_k|0\rangle, \label{eq:odd}
\end{align}
where $|0\rangle$ is the particle vacuum, and thus $\psi(\xi)|0\rangle
\equiv 0$.  The normal number and anomalous densities are in this case
\begin{align}
&n_l(\xi,\xi') = \langle \Phi|\alpha_l\psi^\dagger(\xi')\psi(\xi)\alpha_l^\dagger|\Phi\rangle \nonumber\\
&=\sum_{k \ne l} v_k^*(\xi) v_k(\xi')+u_l(\xi) u^*_l(\xi') ,\label{eq:number1} \\
&\kappa_l(\xi,\xi') = \langle \Phi|\alpha_l\psi(\xi')\psi(\xi)\alpha_l^\dagger|\Phi\rangle \nonumber \\
& =\sum_{k\ne l} v_k^*(\xi)u_k(\xi')+u_l(\xi)v^*_l(\xi'). \label{eq:anomal1}
\end{align} 
Thus the major difference from Eqs.~(\ref{eq:number}-\ref{eq:anomal})
is the absence of the contribution of the chosen quasiparticle state
$l$ in the sum, which is replaced by the ``flipped'' quasiparticle
wavefunction $(u_l(x),v_l(x)) \leftrightarrow
(v^*_l(x),u^*_l(x))$.  This quasiparticle state $l$ is chosen so
as to minimize the total energy of the system with a fixed average odd
fermion number. In the case of an odd-odd nucleus one has to naturally
chose two such quasiparticle states, one for the neutron and the other
for the proton subsystems respectively.

 \section{Self-consistent equations for systems with an odd number of fermions}\label{sec:nuclei}
 
 Here I describe the two chemical potential framework for a polarized
(odd) fermion system, assuming axial symmetry, which apparently 
was first introduced in Ref.~\cite{Bulgac:2007} and in my argumentation here 
I follow the line of reasoning of Ref.~\cite{Bertsch:2009}.  I introduce a new
quantum number $\eta$, the sign of the expectation value of the
single-particle angular momentum operator along the axial symmetry
axis $Oz$ and the corresponding operator
 \begin{align}
 {\cal S}_z=\text{sign}( {j}_z)=\text{sign}\left ({l}_z+\frac{\hbar}{2}\sigma_z\right ).
 \end{align}
 If $Oz$ is the axial symmetry  axis then the quasiparticle wavefunctions are eigenfunctions of 
 ${\cal S}_z$ with eigenvalues 
 \begin{align}
 \eta=\text{sign}(m)=\pm 1
 \end{align}  
where $\hbar m$ are the eigenvalues of $j_z$.
The self consistent equations are:
\begin{align}
\begin{pmatrix}
\text{H}_{\eta \uparrow \uparrow} & H_{ \uparrow \downarrow} & 0 & \Delta \\
H_{ \downarrow \uparrow} & \text{H}_{\eta \downarrow \downarrow}& -\Delta & 0 \\
0 & -\Delta^* & - \text{H}^*_{\eta \uparrow \uparrow} & -H^*_{\eta \uparrow \downarrow} \\
\Delta^* & 0 & -H^*_{\eta \downarrow \uparrow} & -\text{H}^*_{\eta \downarrow \downarrow}
\end{pmatrix}
&\begin{pmatrix}
u_{km,\uparrow} \\
u_{km,\downarrow} \\
v_{km,\uparrow} \\
v_{km,\downarrow}
\end{pmatrix}\nonumber \\
= {\cal H}_{km}\psi_{km}
&\begin{pmatrix}
u_{km,\uparrow}  \\
u_{km,\downarrow} \\
v_{km,\uparrow} \\
v_{km,\downarrow}
\end{pmatrix}, \label{eq:eqp_static}
\end{align}
\begin{align}
&\text{H}_{\eta \sigma,\sigma} = H_{\sigma \sigma} -\mu-\mu_\eta {\cal S}_z -Q,\quad \text{with} \quad  \sigma=\uparrow,  \downarrow, \label{eq:hss}
\end{align}
where $\psi_{km}$ is the four component quasiparticle wave function
and where I suppressed the arguments $({\bm r},\sigma)$ for spatial
and spin coordinates (isospin is not explicitly displayed).  $Q=\sum_l
\lambda_lQ_{l0} $ stands for all other necessary constraints,
including the corresponding Lagrange multipliers, and $k$ stands for
the rest of quantum numbers characterizing the quasiparticle states,
apart from $m$.  I have also used a short hand notation for the
components of the quasiparticle wavefunctions $u_{km,\sigma}
=u_{km}({\bm r},\sigma), \; v_{km,\sigma} =s_{km}({\bm r},\sigma)$,
where $\sigma = \uparrow,\downarrow$.  In this case there are two
chemical potentials $\mu+\eta\mu_\eta=\mu \pm \mu_\eta$.

The normal partial and total number (and other relevant) densities and
also anomalous densities are
\begin{align}
&    n_\eta({\bm r},\sigma,\sigma')= \sum_{E_{km}>0} v_{k m}^*({\bm r},\sigma')v_{k m}({\bm r},\sigma)\delta_{\eta,\text{sign}(m)},\\
&   n({\bm r},\sigma,\sigma')= n_+({\bm r},\sigma,\sigma') +n_-({\bm r},\sigma,\sigma')  ,\\
&   \nu({\bm r})= \sum_{E_{km}>0} v_{km}^*({\bm r},\downarrow)u_{km}({\bm r},\uparrow). \label{eq:dens}
\end{align}
The Hamiltonian $H_{\sigma,\sigma'}({\bm r})$ and the pairing
potential $ \Delta_{\sigma,-\sigma}({\bm r})$ are functional
derivatives of the energy density functional (EDF) ${\cal E}(n, \nu,
\tau, \ldots) $
\begin{align}
& H_{\sigma,\sigma'}({\bm r}) = \frac{ \delta {\cal E}(n, \nu, \tau, \ldots) }{ \delta n({\bm r},\sigma,\sigma') }, \\\
& \Delta_{\sigma,-\sigma}({\bm r})=\frac{\delta {\cal E}(n, \nu, \tau, \ldots)}{\delta \nu({\bm r},\sigma,-\sigma)}.
\end{align}
Note that $\Delta_{\sigma,\sigma}({\bm r})\equiv 0$ in case of
$s$-wave pairing and either proton-proton or neutron-neutron pairing.
The generalization of the formalism to either $s$-wave proton-neutron
pairing or higher partial wave nucleon-nucleon pairing is
straightforward.  The EDF of nuclear systems typically depends on the
sum $n({\bm r},\sigma,\sigma')= n_+({\bm r},\sigma,\sigma') +n_-({\bm
r},\sigma,\sigma')$ alone. In the case of cold atoms the EDF depends
on both $n_\pm({\bm r},\sigma,\sigma')$ separately, as fermions of
various flavors can reside in different external potentials. This
situation is formally equivalent to placing the cold atom system in a
strong ``magnetic'' field, with large ``spin-up and spin-down magnetic
moments,'' see Section~\ref{sec:cold_atoms}.  The two chemical
potentials $\mu\pm\mu_\eta$ are determined from the condition that the
total and partial particle numbers are
\begin{align}
&N=N_+ + N_-, \quad |N_+-N_-|=1,\\ 
&N_\pm = \int d^3{\bm r}\sum_{\sigma=\uparrow,\downarrow} n_\pm({\bm r},\sigma,\sigma).
\end{align}
Notice that the equations for the quasiparticle states with $m>0$
($\eta=+1$) and $m<0$ ($\eta = -1$) respectively have two different
chemical potentials $\mu_{\pm 1}= \mu\pm \mu_\eta$, and in both cases
the eigenvalues $E_{km}$ come also in pairs $(E_{km},-E_{km})$, see
also Section~\ref{sec:cold_atoms}.  

The operator ${\cal S}_z$ can be used in the case of axial symmetry
for a nucleus with octupole deformation, unlike the hermitian
signature operator $i\exp(-i\pi{j}_x/\hbar)$ with eigenvalues $\pm 1$
suggested in Ref.~\cite{Bertsch:2009}.The operator
$R_x=\exp(-i\pi{j}_x/\hbar)=-R_x^\dagger$ is anti-Hermitian, since
$R_x^\dagger R_x=1$ and $R_x^2=-1$ for a single fermion state.  One
can relatively easily use instead of the operator
$i\exp(-i\pi{j}_x/\hbar)$ the so called simplex operator
$i\exp(-i\pi{j}_x/\hbar)P$~\cite{Dobaczewski:2000a}, where $P$ is the
spatial parity operator, if both quadrupole and octupole deformations
are present.  I find the use of the operator ${\cal S}_z$ however much
simpler to implement numerically, particularly if one uses a
coordinate representation of the quasiparticle wave functions on
spatial 3D lattice, see Apendix~\ref{sec:numeric}.  If one uses the
operator ${\cal S}_z$, then quasiparticle states with
$\eta=\text{sign}(m)=\pm 1$ are assigned to particle numbers $N_\pm$
and densities $n_\pm({\bm r},\sigma, \sigma')$ respectively.

Another option would be to use the operator
$i\exp(-i\pi{j}_z/\hbar)=i\exp(-im\pi)=\pm 1$.  In this case
quasiparticle states with $m=+1/2,-3/2,+5/2,\ldots$ are assigned to
particle number $N_+$ and $n_+({\bm r},\sigma, \sigma')$, and states
with $m=-1/2,+3/2,-5/2,\ldots$ are assigned to particle number $N_-$
and $n_-({\bm r},\sigma, \sigma')$, respectively.  Time-reversed
partners are in both cases assigned to different groups, but using
different criteria.  
 
 This ambiguity in assigning the quasiparticle states to either one or
another partial number density in the case of an odd fermion number is
general. One can use any rule to separate quasiparticle states in two
groups, and there is no general prescription on how to assign them to
specific particle number (aka $``N_+$'' and ``$N_-$'') or number
densities. The best solution should always correspond to the lowest
total energy, which might not always favor the strongest pairing
correlations. The many-fermion self-consistent equations never have a
unique solution, even for even (in the case of electrons or cold
atoms) or even-even nuclei, though only one of them is the lowest
total energy. Multiple vacua however, can also correspond to
physically realizable states, separated by strong potential barrier, a
situation which is quite ubiquitous in quantum field theories or
infinite many-body systems, where symmetry is spontaneously broken.
The ambiguities I discuss here for odd fermion systems in this sense
should not come as a surprise, but merely as new examples of such
physically realizable ``ground states.''  However, since in the case
of odd fermion systems the density of low energy levels is relatively
high, any such prescription is to some extent arbitrary, and the true
ground state might emerge as an optimal superposition of many such
quasiparticle vacua, with well defined quantum numbers. One of the
simplest examples is that of a system with spontaneous parity
breaking, when the ground and the first excited states are separated
by an exponentially small energy difference, a phenomenon known in the
literature as parity doubling~\cite{Sheline:1989,Dobaczewski:2005}.
Another possibility is that of shape coexistence, see
e.g. Refs.~\cite{Andreyev:2000,Clement:2016} and references therein,
of which there are many examples of other types too.
 
 There is still another ambiguity. If both quadrupole and octupole
deformations are present in a nucleus with axial symmetry, it is not
clear whether the ground state of the nucleus corresponds to either
$N_+-N_-=\pm 1$, and thus whether the projection of the total angular
momentum ${J}_z(n)=\sum_{n=1}^A{j}_z(n)\neq 0$ is along or
opposite to the fission direction. Such a situation can be
experimentally studied in fission induced by a nucleon transfer from
an impinging projectile to a nucleon state with relatively large
angular momentum~\cite{Ramos:2020}.  The projectile will impact an
angular momenta perpendicular to the reaction plane equal to either
$l_z\approx \pm Rk_F$.  The emergence of the fission fragments emitted
along the axis perpendicular to the reaction plane, might favor the
emission of the either the light or of the heavy fission fragment in
the direction of the impacted angular momentum.  It would very
interesting to see if any asymmetry of the fission fragments
distribution along the direction perpendicular to the reaction plane
exists, as such a phenomenon has not been studied yet to my knowledge.

\begin{figure}[ht]
\includegraphics[width=0.85\columnwidth]{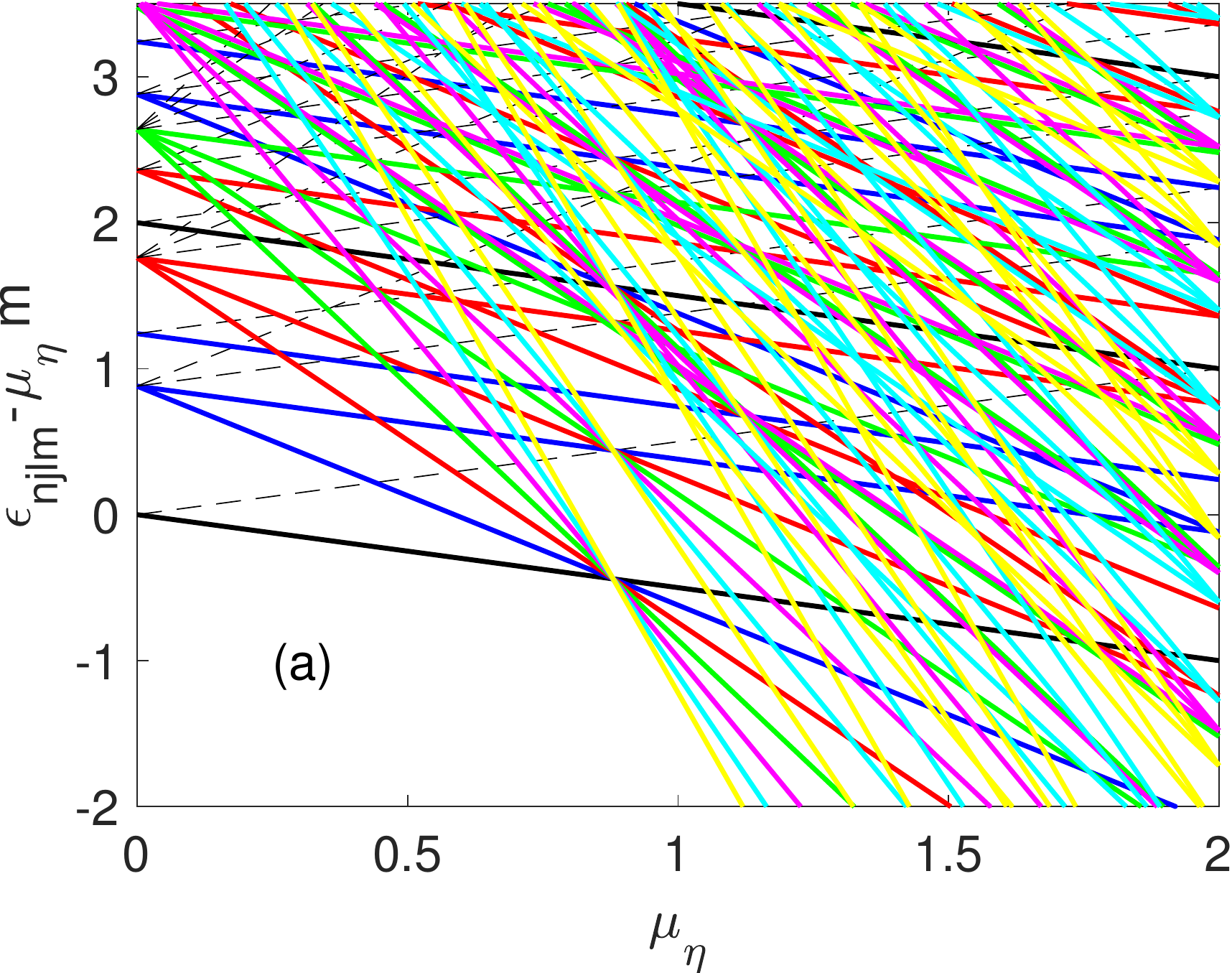}
\includegraphics[width=0.85\columnwidth]{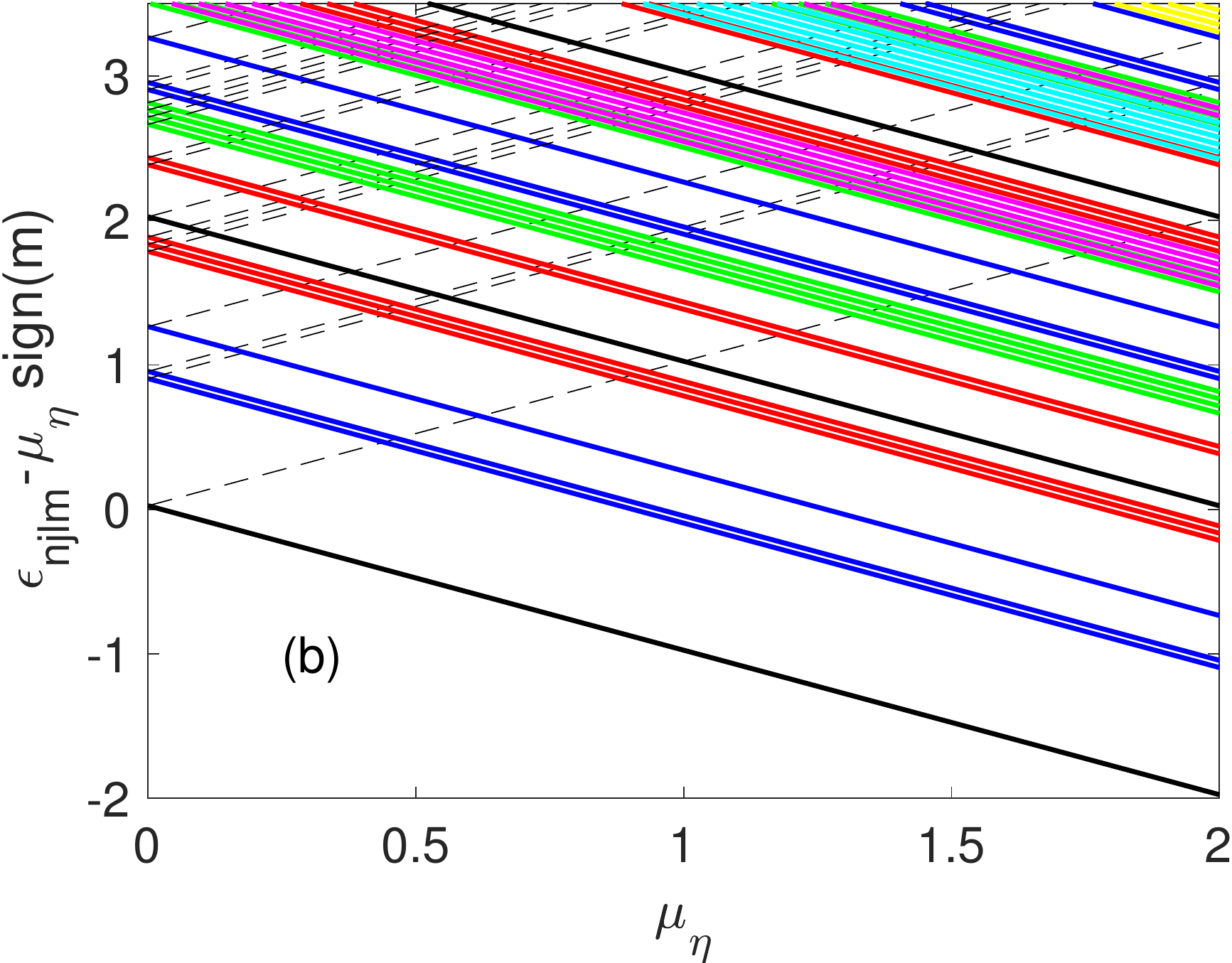}
\caption{\label{fig:hj} (Color online) The colors for down-sloping
quasi-particle levels are black ($l=0$), blue ($l=1$), red ($l=2$),
green ($l=3$), magenta ($l-4$), cyan ($l=5$), and yellow ($l=6$)
respectively.  The corresponding up-sloping levels are always
displayed with black dashed thin lines for the sake of the clarity of
the figure.  In panel (a) are the energy levels for ${H}-\mu_\eta
{j}_x $ and the lower panel for $H-\mu_\eta\;\text{sign}({j}_x).$
Notice that in  panel (b) the states with $j> 1/2$ are multiple
degenerate and for display purposes only have been artificially
perturbed.  For example there are three down-sloping $1d_{5/2}$ levels
corresponding to $m=1/2, 3/2$ and $5/2$.  The single-particle energies 
$\epsilon_{njlm}$, the chemical potential $\mu_\eta$ are in units of
$\hbar \omega$ and the strength of the spin-orbit interaction was 
chosen as $V= 0.12\hbar\omega$ in these examples.}
\end{figure}

In order to illustrate more vividly the difficulties with the
prescriptions described in this section I will apply them to a very
simple case, non-interacting ``neutrons'' in a spherical harmonic
oscillator potential with a constant spin-orbit interaction and no
pairing field:
\begin{align}
H=-\frac{\hbar^2\Delta}{2m}+\frac{m\omega^2 {\bm r}^2}{2} + V{\bm l}\cdot{\bm s}, \label{eq:ho}
\end{align}
as spin-polarization is a property of both normal and superfluid fermionic
systems. Since there is no ``neutron-neutron'' interaction the mean
field is independent of the number of particles, and the ground state
is typically degenerate for even particle numbers.
In Fig.~\ref{fig:hj} I show the 
quasi-particle spectra for the two types of constraints discussed in this section
\begin{align}
{H}-\mu_\eta {j}_z \quad \text{and} \quad {H}-\mu_\eta\;\text{sign}({j}_z).
\end{align}
Since the Hamiltonian has spherical symmetry the choice of constraining axis is arbitrary.
 The two spectra have distinctly different aspects. For small values
of $\mu_\eta$ one can hastily conclude that either type of constraint
apparently leads to desired outcomes. E.g. a system with $N=8$
particles at $\mu_\eta=0$ is a closed shell, but for small finite values
of $\eta$ a particle is promoted from the $1p_{1/2}, m=-1/2$ level to
the $1d_{5/2}$ level.
 
 Since the cranking operator is ${j}_z$ the net result of such a
constraint is imparting the system a net total angular momentum
${J}_z=\sum_{i=1}^N{j}_z(i)$ however, which naturally also gives
raise to a spin-polarization. This is clearly seen in the upper panel of
Fig.~\ref{fig:hj} for $\mu_\eta \approx 1$, when mostly down-sloping
levels with $m>0$ are occupied. The same kind of ground state is
obtained also in the case of constraint $\text{sign}({j}_z).$ In
this case levels with $m>0$ are down-slopping and levels with $m<0$
are up-sloping, all with identical slopes. Nevertheless, the same
undesired characteristic of the system ground state is obtained for a
highly polarized system. In the case of real nucleus, when the depth
of the mean field potential is finite, the radial profiles of the
single-particle wave functions are affected in drastically different
manners for up-sloping (less bound) and down-sloping (more bound)
levels, as a result of a large finite total angular momentum of the
system. The system is polarized because it was forced to have a large
total angular momentum, as opposed to what one might expected to have
a finite total momentum due to the finite popularization of the
system. Basically in either of the prescriptions discussed in this
section and in Ref.~\cite{Bertsch:2009} the roles of the effect and
cause have been inadvertently switched. Since a finite spin-polarization of
a fermionic system corresponds in the mean field approximation to the
breaking of the time-reversal invariance, a finite spin-polarization is
always accompanied by a non-vanishing total angular momentum in {\it
vice versa}. In practice however, one has to clearly distinguish
between what kind of constraint one intends to impose, as different
constraints lead to different outcomes.
 
 \section{The case of fermionic polarized cold atom systems}\label{sec:cold_atoms}
 
\begin{figure}[ht]
\includegraphics[width=1\columnwidth]{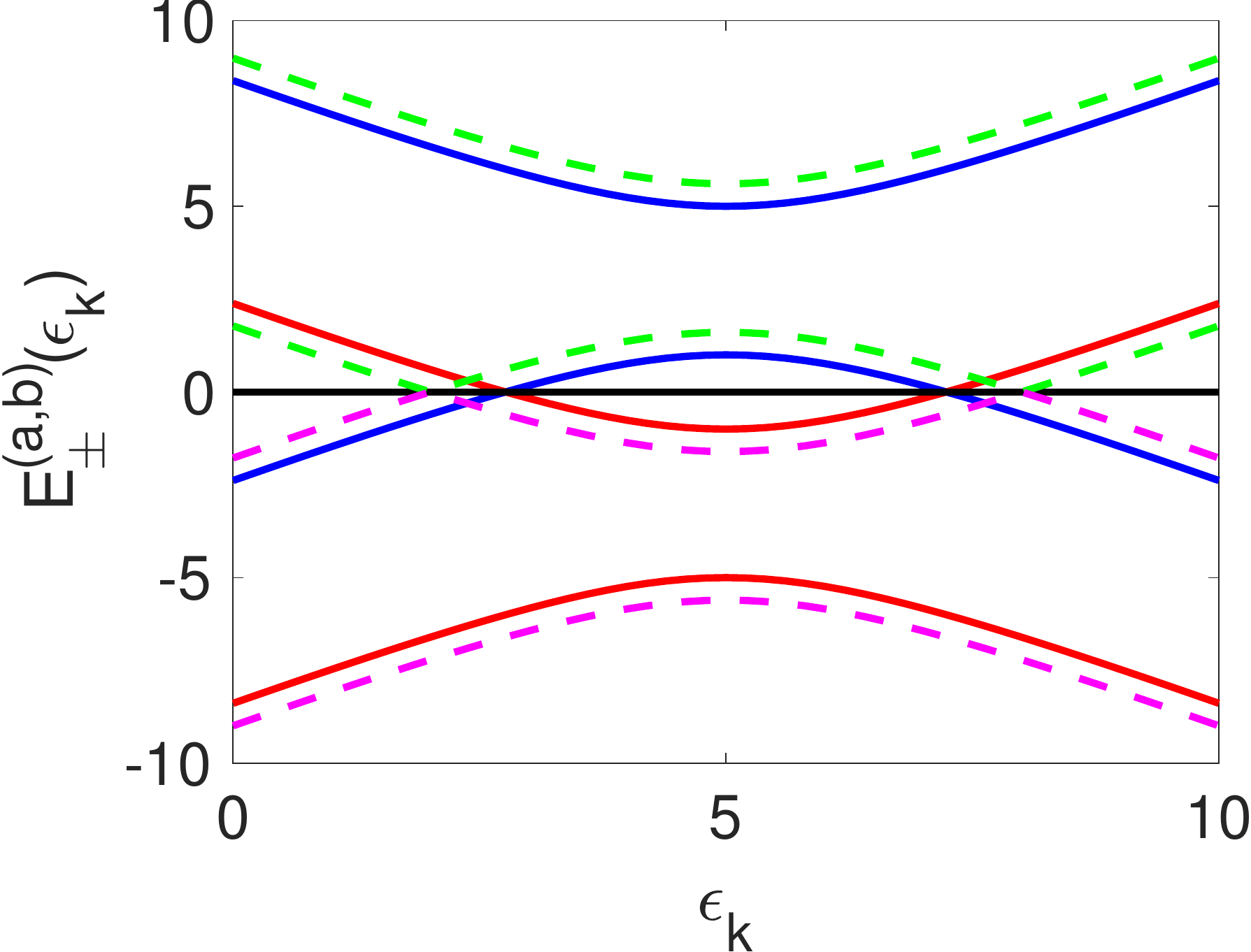}
\caption{\label{fig:Eqp} The quasiparticle spectrum
$E_{k,\pm}^{(a)}$~\eqref{eq:Ea} with blue and
$E_{k,\pm}^{(b)}$~\eqref{eq:Eb} with red lines respectively.  The
densities are constructed from quasiparticle states with
$E_{k,\pm}^{(a,b)}> 0$ only.  For a finite spin-polarization the
quasiparticle states corresponding to the red line with
$E_{k,+}^{(b)}> 0$.  At the same time the fermions of kind $a$ occupy
quasiparticle states corresponding to the blue lines with
$E_{k,\pm}^{(a)}> 0$.  The effect of a non-vanishing mixing between
the two flavors or of a spin-orbit coupling (characteristic to nuclear
systems) on the quasiparticle spectrum is illustrated with dashed
lines. In the case of cold atoms this type of mixing is equivalent to
the creation of a Schr\"odinger cat fermion state between type $a$ and
type $b$ flavors. }
\end{figure}

 In the case of cold atoms one has two flavors of fermions, which I
shall denote with $a$ and $b$, and the Cooper pair is formed between
one fermion $a$ with another fermion $b$. Entangled states, when for
example a type $a$ fermion can coexist with a type $b$ fermion, in a
type of the Schr\"odinger cat single-particle state, have not been
studied yet, neither experimentally nor theoretically to my knowledge.
Only the formation of Copper pairs between a type $a$ fermion and a
type $b$ fermion have been considered so far in the literature. Such a
mixing is formally similar to the spin-orbit coupling of the nucleon
motion in nuclei, and it will be illustrated qualitatively in
Fig.~\ref{fig:Eqp}.  In the absence of such mixing the mean field
equations
read~\cite{Bulgac:2007,Sensarma:2007,Bulgac:2008,Bulgac:2011a,aslda}:
 \begin{align}
&\begin{pmatrix}
H_a -\mu_a & 0 & 0 & \Delta \\
0 & H_b-\mu_b & -\Delta & 0 \\
0 & -\Delta^* &  -H^*_a  +\mu_a& 0 \\
\Delta^* & 0 & 0 & -H^*_b +\mu_b
\end{pmatrix}
\begin{pmatrix}
u_{k}^{(a)} \\
u_{k}^{(b)} \\
v_{k}^{(a)} \\
v_{k}^{(b)}
\end{pmatrix} \nonumber\\
&= {\cal H}
\begin{pmatrix}
u_{k}^{(a)}  \\
u_{k}^{(b)} \\
v_{k}^{(a)} \\
v_{k}^{(b)}
\end{pmatrix}
= E_{k}
\begin{pmatrix}
u_{k}^{(a)}  \\
u_{k}^{(b)} \\
v_{k}^{(a)} \\
v_{k}^{(b)}
\end{pmatrix}. \label{eq:cold}
\end{align}
where the two chemical potentials $\mu_{a,b}$ are chosen by fixing the
particle numbers
\begin{align}
N_{a,b}=\sum_{E_k>0}  \int d^3{\bm r} |v_k^{(a,b)}({\bm r})|^2.
\end{align}

These equations obviously decouple and since in case of cold atoms one
typically has $H_{a,b}=H_{a,b}^*$, the equations simplify.  By
introducing
\begin{align}
H_\pm = \frac{H_a\pm H_b}{2}
\end{align}
these equations can be re-written as 
\begin{align}
&\begin{pmatrix}
H_+ -\mu & \Delta \\
\Delta^* & -H_+ +\mu
\end{pmatrix}
\begin{pmatrix}
u_{k}^{(a)} \\
v_{k}^{(b)}
\end{pmatrix}\label{eq:cold3}\\
&= 
\begin{pmatrix}
E_k^{(a)} -(H_- -\mu_\eta) & 0 \\
0 & E_k^{(a)} -(H_- -\mu_\eta)
\end{pmatrix}
\begin{pmatrix}
u_{k}^{(a)}  \\
v_{k-}^{(b)}
\end{pmatrix},\nonumber \\
&\begin{pmatrix}
H_+ -\mu & -\Delta \\
-\Delta^* & -H_+ +\mu
\end{pmatrix}
\begin{pmatrix}
u_{k}^{(b)} \\
v_{k}^{(a)}
\end{pmatrix} \label{eq:cold4}\\
&=
\begin{pmatrix}
E_k^{(b)} +(H_- -\mu_\eta) & 0 \\
0 & E_k^{(b)} +(H_- -\mu_\eta)
\end{pmatrix}
\begin{pmatrix}
u_{k}^{(b)}  \\
v_{k}^{(a)}
\end{pmatrix},
\nonumber
\end{align}
and now one can disentangle the different roles operators $H_+-\mu$
and $H_- -\mu_\eta$ play on acting on quasiparticle wave functions.
The chemical potentials $\mu$ and $\mu_\eta$ are defined in a similar
manner
\begin{align}
\mu = \frac{\mu_a+\mu_b}{2},\quad \mu_\eta=\frac{\mu_a-\mu_b}{2}.
\end{align}
The self-consistent equations in the nuclear case can be brought to a similar form.
Assuming that $H_{a,b}=\varepsilon_{a,b}$ and $\Delta$ are diagonal one can show that
\begin{align}
& E_{k,\pm}^{(a)} = +e_- \pm \sqrt{ \epsilon_k^2+|\Delta |^2 }, \label{eq:Ea} \\
&E_{k,\pm}^{(b)} = -e_-  \pm \sqrt{ \epsilon_k^2+|\Delta |^2 },\label{eq:Eb}\\
&e_-= \frac{\varepsilon_a-\varepsilon_b}{2} -\mu_\eta, \quad 
  \epsilon_k=\frac{\varepsilon_a+\varepsilon_b}{2} -\mu.
 \end{align}
Clearly all eigenvalues come in pairs $E_{k,\pm}^{(a)} =
-E_{k,\mp}^{(b)}$.  For each eigenvector $(u^{(a)}_k,v^{(b)}_k)$ and
corresponding eigenvalue $E_{k,\pm}^{(a)}$ of Eq.~\eqref{eq:cold3} the
Eq.~\eqref{eq:cold4} has a corresponding eigenvector $
(u^{(b)}_k,v^{(b)}_k)= (v^{(a)}_k,u^{(b)}_k)^*$ and a corresponding
eigenvalue $E_{k,\mp}^{(b)}=-E_{k,\pm}^{(a)}$.

As branches of the quasiparticle spectrum are displaced in opposite
directions, when part of the lower branch $E^{(a)}_{k,-}$ becomes
positive (with blue in Fig.\ref{fig:Eqp}) and the upper branch
$E^{(b)}_{k,+}$ becomes negative (with red in Fig.\ref{fig:Eqp}), the
roles of the components of the Bogoliubov quasiparticles change
exactly as discussed in Section~\ref{sec:general}, see
Eqs.~(\ref{eq:number1}, \ref{eq:anomal1}). If $N_a-N_b$ quasiparticle
energies $E_{k,-}^{(a)}$ change their signs, then the new
quasiparticle vacuum corresponds to
$\prod_{l=1}^{N_a-N_b}\alpha_{\mu_l}^\dagger|\Phi\rangle$ and a
particle parity $(-1)^{N_a-N_b}$, where $|\Phi\rangle$ is a total
fermion function for an unpolarized system, and $\mu_l$ are the
corresponding quantum numbers of the positive $E_{k,-}^{(a)}$
quasiparticle states.

When the branches of $E_{k,-}^{(a)}$ and $E_{k,+}^{(b)}$ cross zero,
some of the quasiparticles energies could vanish identically, as in
the case of bound states on a superfluid vortex line first discussed
by \textcite{Caroli:1964} and have a character similar to Majorana
particles. In such a case the fermion system is technically a
topological one, characterized by a Chern number associated with the
Berry connection and curvature~\cite{Thouless:1982}.

\section{My favorite two chemical potentials framework for a polarized superfluid Fermi system} \label{sec:best}

Likely the best option is to formulate the two chemical potentials
framework for nuclei along the same lines as for the scheme suggested
for cold atoms, see
Refs.~\cite{Bulgac:2007,Sensarma:2007,Bulgac:2008,Bulgac:2011a} and
Section~\ref{sec:cold_atoms}. The main difference between nuclei and
cold atom systems is in the presence of the spin-orbit interaction in
nuclei and the need to introduce the total single-particle angular
momentum ${\bm j}={\bm l}+{\bm s}$, where ${\bm l}$ is
the orbital angular momentum and ${\bm s}= \hbar{\bm
\sigma}/2$ is the nucleon spin, and the spin-orbit interaction, which
mixes the spin-up and spin-down states, which is qualitatively
different situation from the cold atom case discussed in the previous
Section~\ref{sec:cold_atoms}. (
 
 The main problem with either the method suggested by
\textcite{Bertsch:2009} and briefly described in
Section~\ref{sec:nuclei} or the ``improvement'' suggested by me in the
same section is that either the constraining operators ${\cal
S}_{x,z}$ lead to polarization, exactly as any rotation with a finite
frequency would do or a strong magnetic field also induce.  The goal
is not to bring the nucleus into rotation, and thus excite it, but
rather to keep the odd or odd-odd nucleus in its ground state at 
a given spin-polarization in the
mean field approximation. The question arises then, what is the most
appropriate constraining operator?
 
 When $\exp\left( -i2\pi {\bm
j}\cdot{\bm n}/\hbar\right )$ is acting on a $1/2$-spinor wave function
it changes its sign.
The rotation operator  can be
factorized, since $[{{\bm l},{\bm s}}]=0$
 \begin{align}
 \exp\left ( -i\alpha \frac{{\bm j}\cdot{\bm n}}{\hbar}\right ) =
 \exp\left ( -i\alpha \frac{{\bm l}\cdot{\bm n}}{\hbar} \right )
 \exp \left ( -i\alpha \frac{{\bm s}\cdot{\bm n}}{\hbar} \right ),
 \end{align}
 where ${\bm n}=(\sin\beta\cos\gamma,\sin\beta\sin\gamma,\cos\beta)$ is an
arbitrary unit vector.  If one considers now separately the action of
$\exp( -i2\pi {\bm l}\cdot{\bm n}/\hbar)$ on any component of the
spinor, that component does not change sign. However, when acting with
$ \exp \left ( -i2\pi {\bm s}\cdot{\bm n}/\hbar\right )$ alone on
the $1/2$-spinor wave function, the spatial part of the wave function
is obviously unaffected, but the whole spinor wave function changes
sign. That allows us to define the operator
 \begin{align}
 {\cal P}_N(\alpha) = \prod_{k=1}^N  \exp \left [ -i\alpha\frac{ {\bm s}(k)\cdot{\bm n}}{\hbar}\right ],
\end{align}
where the product runs over all particles. 
\footnote{ Note that the spin operator ${\bm s}$ in this expression 
can be formed from arbitrary Pauli matrices $\tilde{ {\bm \sigma} }$, 
which are defined through arbitrary angles $\kappa,\chi,\zeta$, 
see \eqref{eq:Pauli}.}
The action of this operator leads to an apparently new and simple
representation of the particle parity operator for an $N$-fermion
system
\begin{align} 
{\cal P}_N(2\pi)|\Phi\rangle =(-1)^N|\Phi\rangle .
\end{align}
 It is particularly easy to check that $\langle\Phi | {\cal
P}_N(2\pi)|\Phi\rangle \equiv (-1)^N$ in the case of a Slater
determinant.  Since the spin direction can be chosen arbitrarily, this
representation of the particle projection operator is clearly not
unique and one can use ${\bm n}=(0,0,1)$.  Exactly as in the case
described in Section~\ref{sec:nuclei} the single-particle operator
${\cal P}_1(\alpha) = \exp(-i\alpha{\bm s}\cdot {n}/\hbar) $ has the
obvious properties
\begin{align}
&{\cal P}^\dagger_1(\alpha){\cal P}_1(\alpha)=1, \quad {\cal P}_1^2(\pi)={\cal P}_1(2\pi) =-1
\end{align} 
and all the argumentation presented in Section~\ref{sec:nuclei} and in
Ref.~\cite{Bertsch:2009} follows, however, without some of the
limitations discussed in Section~\ref{sec:nuclei} and other
limitations discussed in Appendix~\ref{sec:numeric}.

A polarized Fermi system is spin polarized, similarly to the well
studied cold atom systems~\cite{Bulgac:2007,Bulgac:2008,Sensarma:2007, 
Bulgac:2011a,Wlazlowski:2018,Tuzemen:2019,Magierski:2019}
or magnetized electron systems for example.
If there are no non-vanishing time-odd external or components of the
mean field one can show that the ground state of an even-even nucleus
has a vanishing total spin $\langle \Phi | \sum_{k=1}^N {\bm
s}(k)|\Phi \rangle\equiv 0$~\cite{Vautherin:1972}.  Hence, 
following the same argumentation presented in Ref.~\cite{Bertsch:2009} 
and in Section~\ref{sec:nuclei},
for a polarized fermion superfluid system Eqs.~(\ref{eq:eqp_static},
\ref{eq:hss}) can be rewritten as follows:
\begin{align}
\begin{pmatrix}
\text{H}_{\eta \uparrow \uparrow} & H_{ \uparrow \downarrow} & 0 & \Delta \\
H_{\downarrow \uparrow} & \text{H}_{\eta \downarrow \downarrow}& -\Delta & 0 \\
0 & -\Delta^* & - \text{H}^*_{\eta \uparrow \uparrow} & -H^*_{\eta \uparrow \downarrow} \\
\Delta^* & 0 & -H^*_{\eta \downarrow \uparrow} & -\text{H}^*_{\eta \downarrow \downarrow}
\end{pmatrix}
&\begin{pmatrix}
u_{k,\uparrow} \\
u_{k,\downarrow} \\
v_{k,\uparrow} \\
v_{k,\downarrow}
\end{pmatrix}\nonumber \\
={\cal H}
\begin{pmatrix}
u_{k,\uparrow}  \\
u_{k,\downarrow} \\
v_{k,\uparrow} \\
v_{k,\downarrow}
\end{pmatrix}
= E_{k}
&\begin{pmatrix}
u_{k,\uparrow}  \\
u_{k,\downarrow} \\
v_{k,\uparrow} \\
v_{k,\downarrow}
\end{pmatrix}, \label{eq:nuclear}
\end{align}
\begin{align}
&\text{H}_{\eta \sigma,\sigma} = H_{\sigma \sigma} -\mu-\eta \mu_\eta
-Q,\label{eq:mu_nuc}
\end{align}
 with $\eta = (\sigma_z)_{\sigma,\sigma}=\pm 1$ for
$\sigma=\uparrow,\downarrow$ respectively.  Here I have replaced
either the operator ${\cal S}_z$ introduced in
Section~\ref{sec:nuclei}, or the operator $i\exp(-i\pi{j}_x/\hbar)$
introduced by ~\textcite{Bertsch:2009}, or the alternative simplex
operator $i\exp(-i\pi{j}_x/\hbar)P$~\cite{Dobaczewski:2000a}, both of
them discussed in Section~\ref{sec:nuclei}, with the much simpler
Hermitian operator $i{\cal P}_1(\pi)=i\exp( -i\pi\sigma_z/2)\equiv
\sigma_z$.  Then the number densities, the particle numbers, and the
anomalous density are determined as follows
\begin{align}
&   n_{\sigma}({\bm r})= \sum_{E_k>0} v_{k }^*({\bm r},\sigma)v_{k }({\bm r},\sigma),\quad \sigma=\uparrow,\downarrow, \label{eq:n_ud}\\
&   n({\bm r})= n_{\uparrow}({\bm r})+  n_\downarrow({\bm r}),\label{eq:nnn} \\
& N=N_\uparrow + N_\downarrow, \quad  N_{\uparrow,\downarrow} = \int d^3{\bm r} n_{\uparrow,\downarrow}({\bm r}), \label{eq:N_ud}\\
&   \nu({\bm r})= \sum_{E_{k}>0} v_{k}^*({\bm r},\downarrow)u_{k}({\bm r},\uparrow). 
\end{align}

 The chemical potentials $\mu_{\uparrow,\downarrow}=\mu\pm \mu_\eta$ 
 are chosen so as fix  both the total particle  
 number $N$ and the degree of spin-polarization $N_\uparrow-N_\downarrow$.
Since the chemical potential $\mu_\eta$ enters in the self-consistent
equations Eq.~\eqref{eq:nuclear} as $\mu_\eta\sigma_z$ the
single-particle angular momentum ${ j}_z$ still commutes with the
quasiparticle Hamiltonian ${\cal H}$ for an axially symmetric
nucleus. However, a spin polarized odd or odd-odd nucleus strictly
speaking cannot be strictly spherical anymore in the mean field
approximation, as $[{\bm j}^2,\sigma_z]\ne 0$.

Naturally, the same procedure can be used for a nucleus without pairing
correlations.  The chemical potential $\eta\mu_\eta$ act as a
fictitious magnetic field $ B|\mu_N|=\mu_\eta$ in case of neutral
particles.  Consider for example the case of $N$ fermions occupying a
set of (ordered) single-particle levels $\varepsilon_k$, all with
Kramers degeneracies.  By applying the ``external field'' $\mu_\eta
> (\varepsilon_{N}-\varepsilon_1)/2$ the system is completely
polarized and partially polarized for smaller values of $\mu_\eta$.

By placing a nucleus in an external magnetic field of arbitrary
magnitude one can also obtain non-vanishing spin-polarizations. However,
protons will respond more strongly to a magnetic field, due to their
finite orbital magnetic moment.  Neutrons and protons cannot be
controlled independently in the present framework, as various applications might
require, for example when there is a multi-nucleon transfer reaction
and one can populate various proton and neutron single-particle
states.

The quasiparticle spectrum is not affected qualitatively by the
presence or absence of the spin-orbit interaction, see
Fig.~\eqref{fig:Eqp}.  Notice that in these formulas I have dropped
the additional subscript $m$ for the energies and the quasiparticle
wave functions, as there is no need for singling out $m$. Various
branches of the quasiparticle spectrum are shifted upwards and
downwards, see Fig.~\eqref{fig:Eqp}, and the quasiparticle wave
functions corresponding to the lowest energies are automatically
``flipped'' when these quasiparticle energies change sign, see the
discussion in previous Sections~\ref{sec:general}, \ref{sec:nuclei},
and \ref{sec:cold_atoms}.

In the traditional approach used for odd fermion systems, see
Refs.~\cite{Ring:2004,Dobaczewski:1997,Schunck:2017,Bertsch:2009a} and
Section~\ref{sec:general} the quantum numbers for the singled out
quasiparticle state $l$, see Eq.~\eqref{eq:odd}, are {\it a priori}
unknown. In order to determine the ground state one needs to perform
many simulations with various choices of quantum numbers for
quasiparticle state $l$~\cite{Schunck:2017,Bertsch:2009a}. In this
latest formulation, c.f.  Eqs.~\eqref{eq:nuclear}, one needs only to
specify the degree of spin-polarization $ N_\uparrow - N_\downarrow $ of
the system only. $| N_\uparrow - N_\downarrow |$ could be any integer
in principle and therefore in this framework one can generate two or
more quasiparticle excited states as well when needed.  One can even
consider $N_\uparrow - N_\downarrow$ as any real number and consider
fractional total particle numbers~\cite{Dreizler:1990lr},
similarly to what has been done in the case of cold atom systems for
arbitrary spin-polarizations, and where the agreement of the EDF approach
with {\it ab initio} quantum Monte Carlo calculations of inhomogeneous
systems and also with experiments was
excellent~\cite{Bulgac:2007,Bulgac:2008,Bulgac:2011a,
Bulgac:2014,Wlazlowski:2015}.  $N_\uparrow-N_\downarrow$ can take any
value in both even and odd nuclei, a situation which might be very
useful when analyzing nuclei in extremely strong magnetic fields,
e.g. in magnetars.

\begin{figure}[h]
\includegraphics[width=0.99\columnwidth]{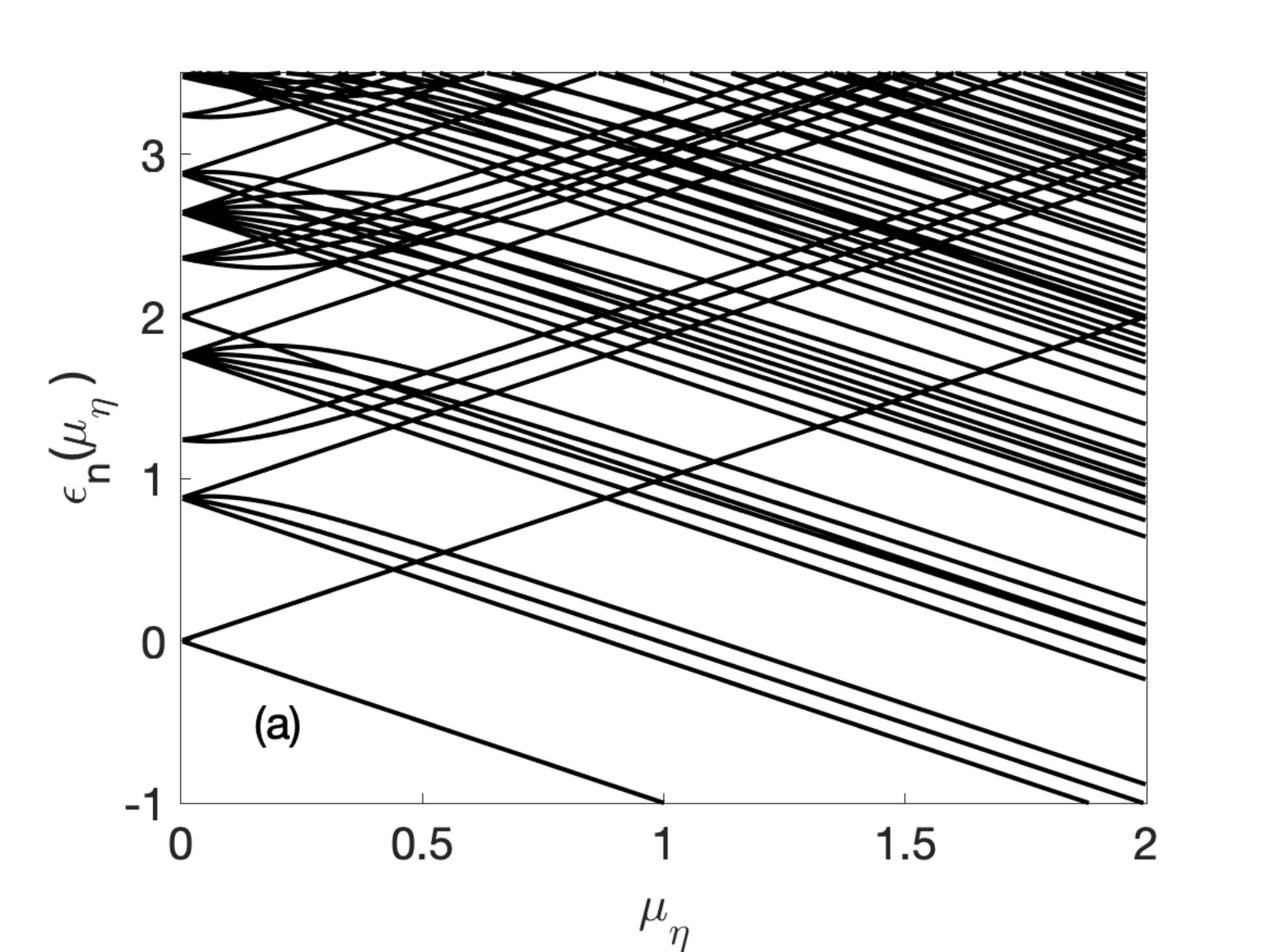}
\includegraphics[width=0.83\columnwidth]{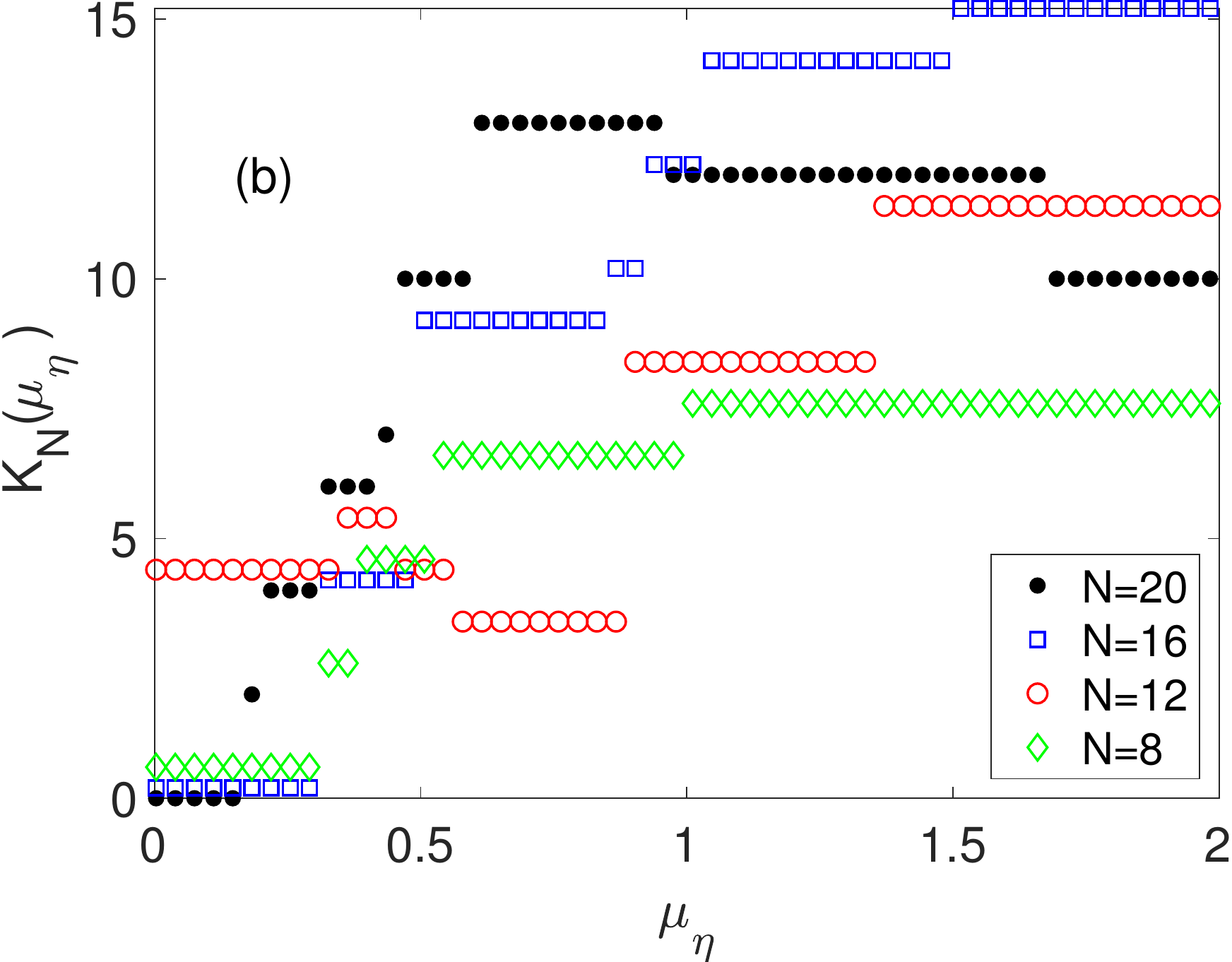}
\caption{\label{fig:en} (Color online) 
The eigenvalues $\epsilon(\eta)$ of the constrained Hamiltonian 
$H-\mu_\eta\sigma_z$, with $h$ given by Eq.~\eqref{eq:ho}, are shown 
in panel (a). The corresponding projection of the total angular momentum 
along the symmetry axis
$K_N(\mu_\eta)= \langle \Psi(\mu_\eta) |\sum_{i=1}^N {j}_x(i) |\Psi(\mu_\eta)\rangle$ 
for a ``neutron'' with $N=8, 12, 16$ and 20 particles  system is shown 
in panel (b). $\Psi(\mu_\eta)$ is the ground state Slater determinant. 
For visual clarity $K_N(\mu_\eta)$ for different $N$ have 
been slightly shifted in the vertical direction for better visualization. 
The energies, $\mu_\eta$,  
and the angular momentum are in units of $\hbar\omega$ and $\hbar$ respectively.  
$N=8$ and 20 are magical numbers, and $N=16$ as zero spin-polarization corresponds to a 
complete filling of the $1s_{1/2},1p_{3/2},1p_{1/2},1d_{5/2}, 2s_{1/2}$ levels and at 
$\mu_\eta\equiv 0$ have  $K_N(0)=0$ and also $\Sigma_N(0)\equiv 0$ as expected, see Fig~\ref{fig:sigma}.  }
\end{figure}

I will present now a second argument in favor of the type of constraint discussed in this section. 
For any nucleus one can consider the generalized number density for
one kind of nucleons
\begin{align}
n_{\sigma,\sigma'}({\bm r}) =\sum_{E_k>0}v_k({\bm r},\sigma)v_k^*({\bm r},\sigma').
\end{align}
\textcite{Vautherin:1972} observed that time-reversal invariance
implies that for an even-even nucleus in the ground state
\begin{align}
&n_{\sigma,\sigma'}({\bm r}) \equiv \frac{1}{2}\delta_{\sigma,\sigma'} n({\bm r}),\\
&\sum_{\sigma,\sigma'} n_{\sigma,\sigma'}({\bm r}) {\bm \sigma}_{\sigma',\sigma}\equiv 0.
\end{align}
For an odd-odd or odd nucleus that is not true anymore, as clearly that also holds true for excited states
of an even-even nucleus with a finite total spin-polarization, when
\begin{align}
{\bm \Sigma}_N =\int d^3r \sum_{\sigma,\sigma'} n_{\sigma,\sigma'}({\bm r}) {\bm \sigma}_{\sigma',\sigma} \ne 0 \label{eq:Sigma}
\end{align}
is non-vanishing.  Obviously 
\begin{align}
\int d^3r \sum_{\sigma} n_{\sigma,\sigma}({\bm r})=N
\end{align} 
is the total number of either neutrons or protons.  Without any loss
of generality one can choose ${\bm \Sigma}= (0,0,\Sigma)$ and then one
can easily see that for any nucleus
\begin{align}
N_{\sigma}=  \int d^3r n_{\sigma,\sigma}({\bm r}), \quad \textrm{for}\quad \sigma=\uparrow,\downarrow,
\end{align}
in agreement with Eqs.~(\ref{eq:n_ud},\ref{eq:nnn},\ref{eq:N_ud}). 
Obviously the spin-polarization $\Sigma_N$ has always  values in a finite interval
\begin{align}
\Sigma_N\in [ -N, N]
\end{align}
and in case of nuclear systems one can consider only non-negative values for $\Sigma_N$.

\begin{figure}[h]
\includegraphics[width=0.85\columnwidth]{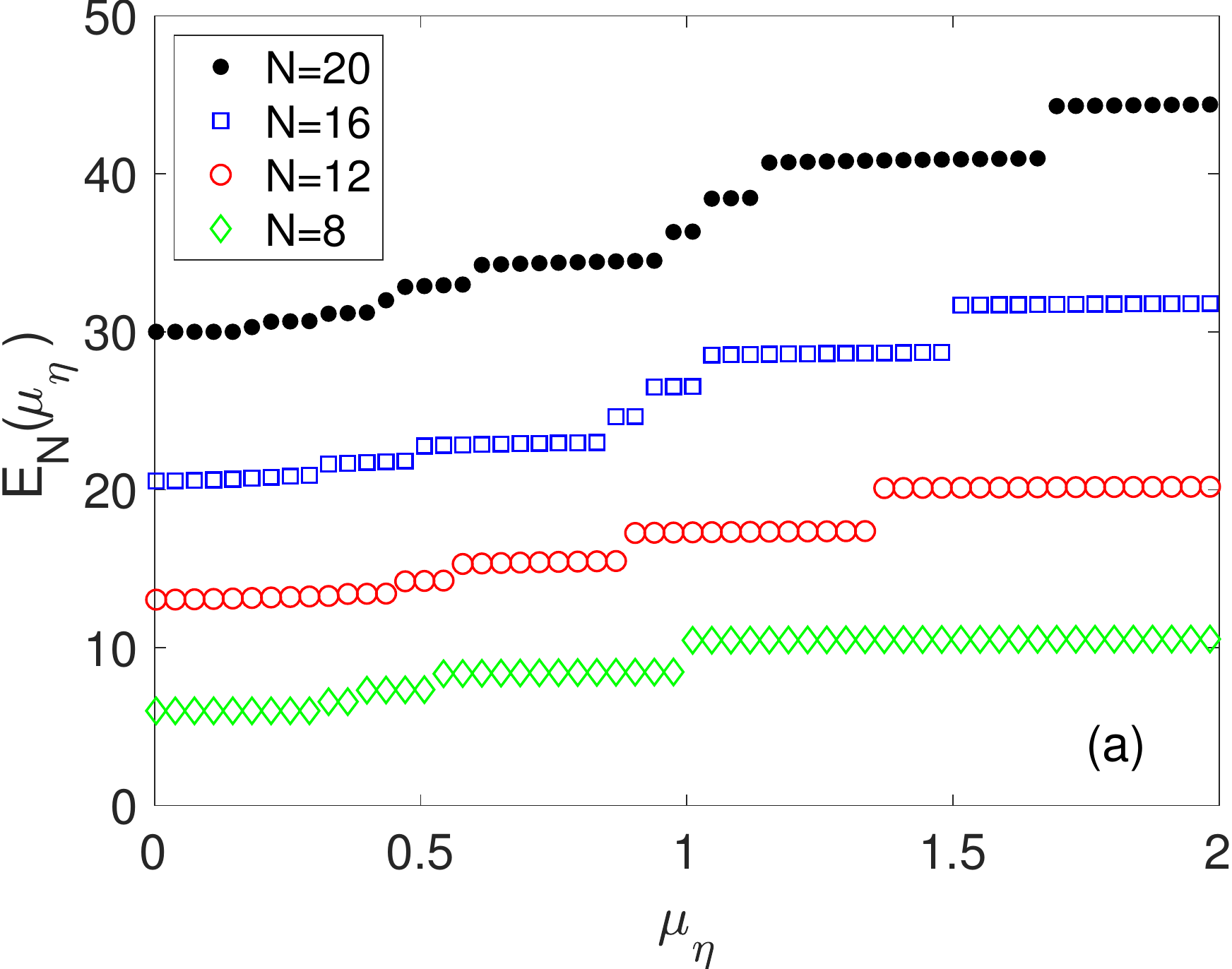}
\includegraphics[width=0.85\columnwidth]{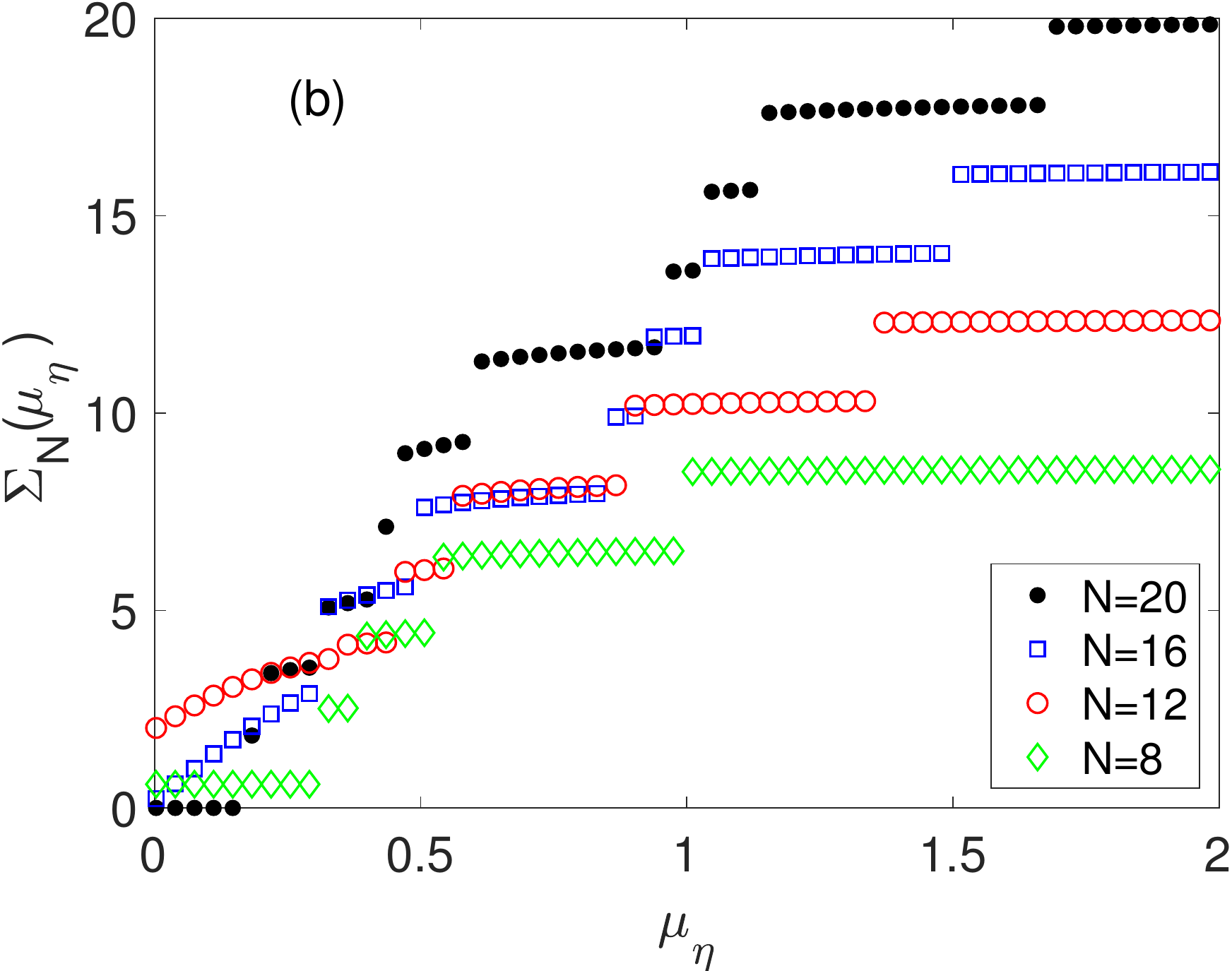}
\caption{\label{fig:sigma} (Color online) The total energy $E_N(\mu_\eta)$
of a normal polarized $N=8, 12, 16$ and $20$ non-interacting
``neutron'' system with Hamiltonian~\eqref{eq:ho} is shown in 
panel (a) as a function of $\mu_\eta$ and the total spin-polarization ${\bm
\Sigma}_N (\mu_\eta)$, see Eq.~\eqref{eq:Sigma}, is shown in 
panel (b). For visual clarity $\Sigma_N(\mu_\eta)$ for different $N$ have 
been slightly shifted in the vertical direction for better visualization. 
Energy and $\mu_\eta$ are in units of $\hbar\omega$. }
\end{figure}

\begin{figure}[ht]
\includegraphics[width=0.85\columnwidth]{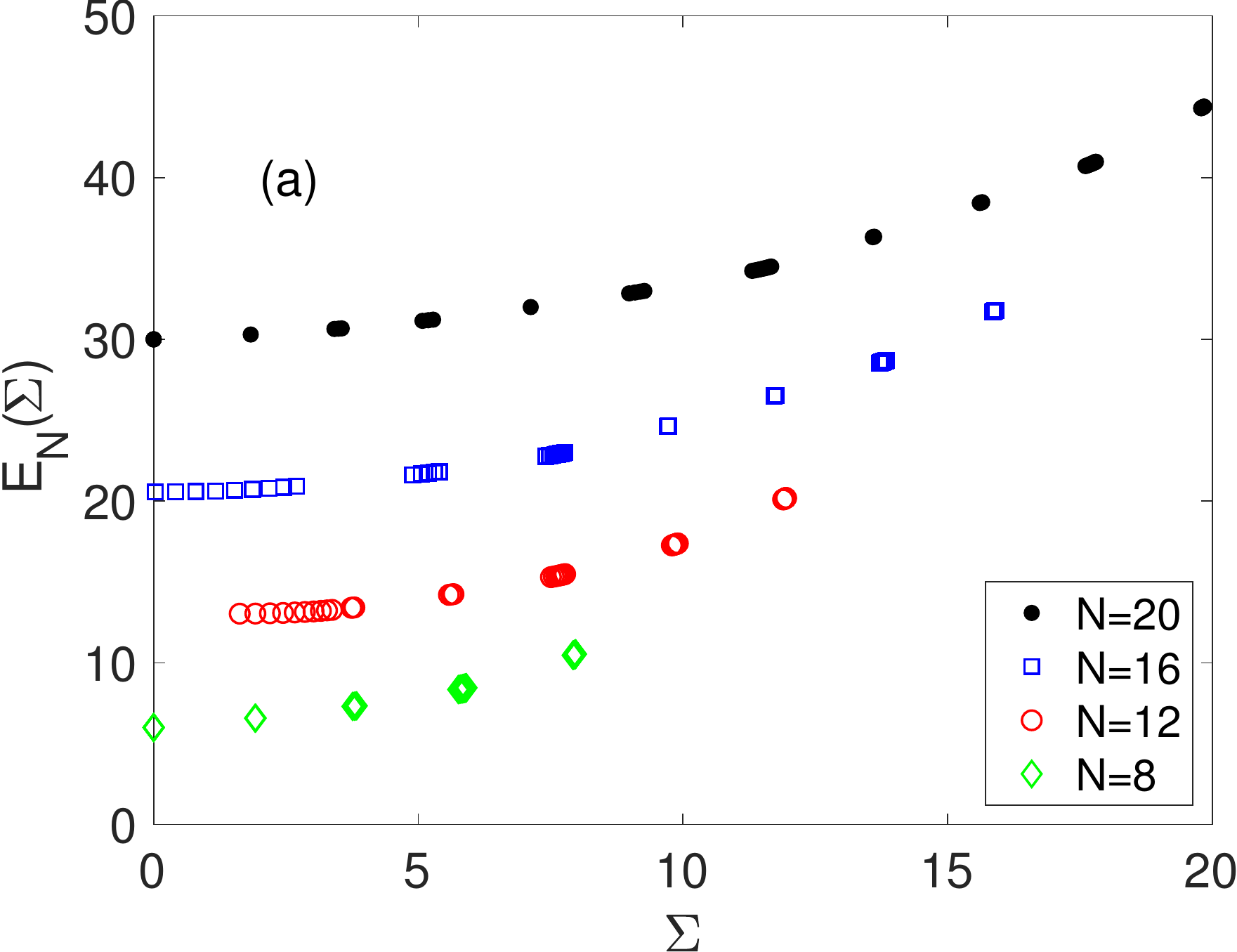}
\includegraphics[width=0.85\columnwidth]{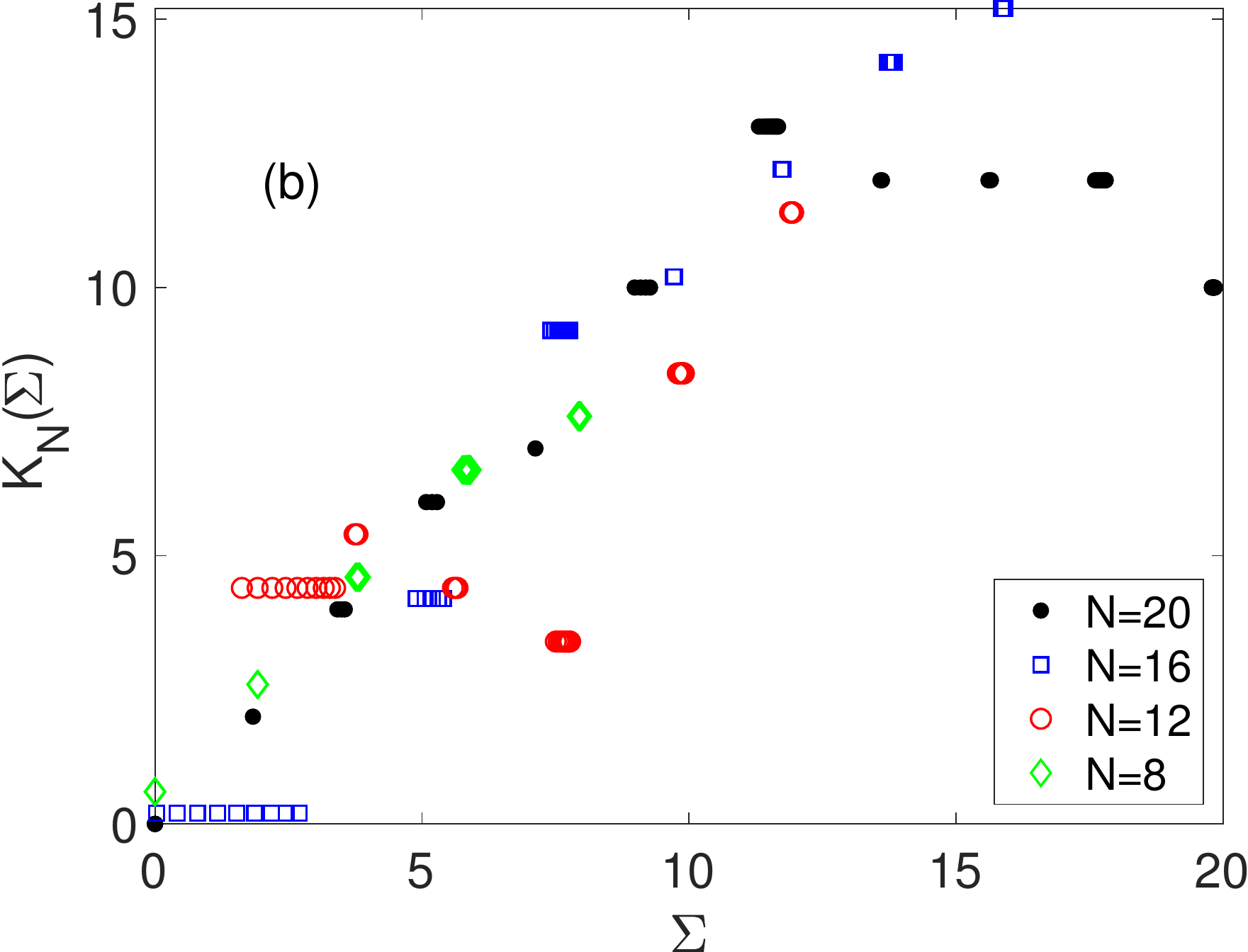}
\includegraphics[width=0.82\columnwidth]{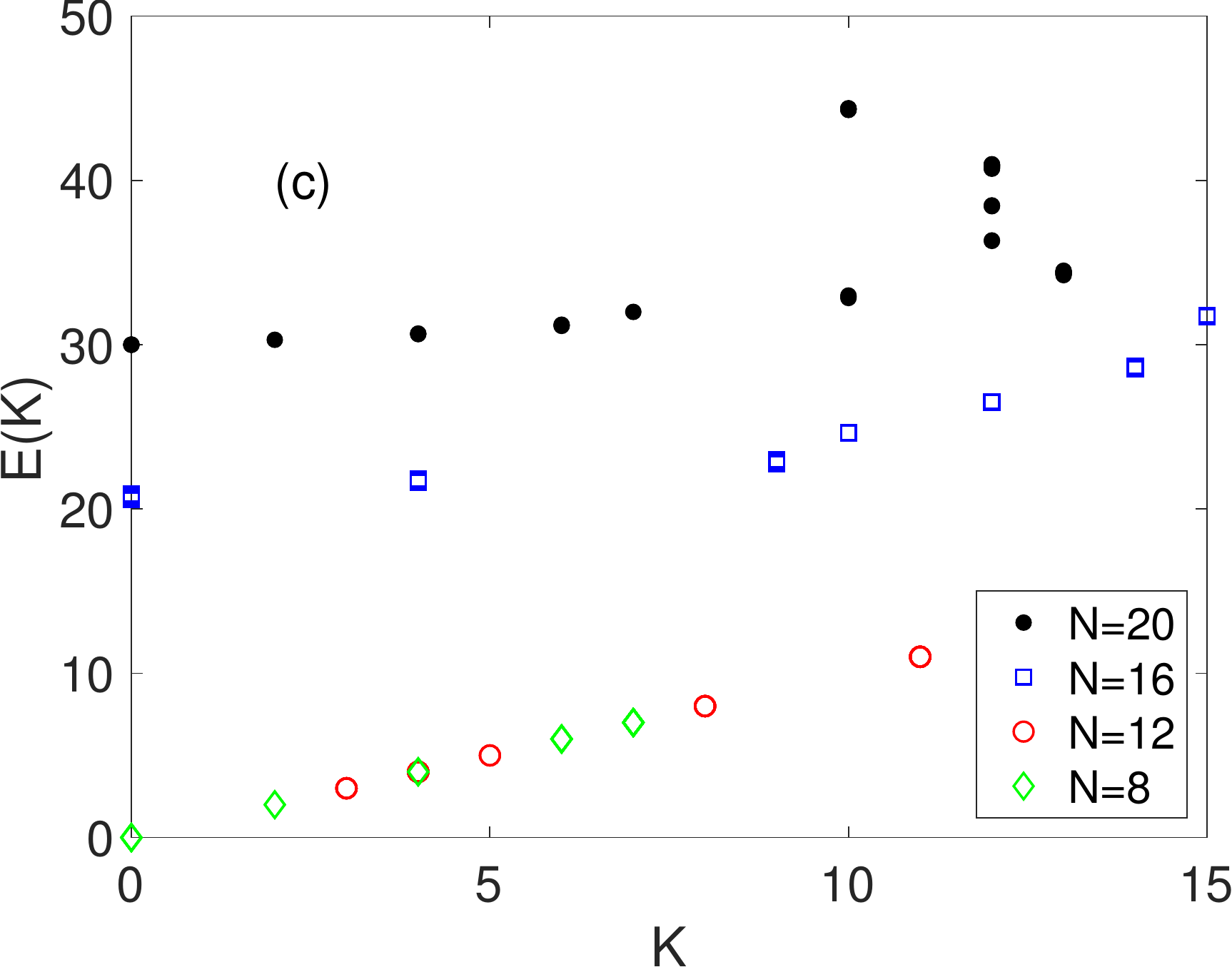}
\caption{\label{fig:ES} (Color online) The total energy $E_N(\Sigma)$ in panel (a), 
the projection of the total angular momentum $K_N(\Sigma)$ along the symmetry axis, 
and the `yrast lines''  $E(K)$ {\it versus} $K$, all  for $N=8, 12, 16$, and 20 are shown in panel (c). 
Energy is in units of $\hbar\omega$ and the angular momentum in units of $\hbar$.
For visual clarity $K_N(\mu_\eta)$ for different $N$ have 
been slightly shifted in the vertical direction for better visualization.  }
\end{figure}

I will compare now the results of imposing this type of constraint on
the system discussed at the end of Section~\eqref{sec:nuclei}, see
Eq.~\eqref{eq:ho}.  Figs.~\ref{fig:en} and \ref{fig:sigma} summarize
the results for quasi-particle energies $\epsilon_n(\mu_\eta)$  for the constrained 
Hamiltonian $H-\mu_\eta\sigma_z$ and for the projection of the total 
angular momentum along the symmetry axis, energy, and spin-polarization  respectively 
$K_N(\mu_\eta), E_N(\mu_\eta)$ and $\Sigma_N(\mu_\eta)$
for the total particle numbers $N=8, 12, 16$ and 20.
In Fig.~\ref{fig:ES} for these
values of $N$ I display the dependence of $E_N(\Sigma)$, $K_N(\Sigma)$, and the ``yrast line'' 
$E(K)$ {\it versus } $K$.  
The total angular momentum, the
total energy, and the total spin-polarization change by jumps at each
quasi-particle level crossing.  The total spin-polarization
$\Sigma_z(\mu_\eta)$ changes in jumps of $\approx 2$ as expected at each
value of $\eta$ where a quasi-particle level crossing occurs, see Fig.~\ref{fig:en}.  There
is a glaring difference between the maximum total angular momentum
obtained at maximum spin-polarization in Fig.~\ref{fig:hj} for the
constraints $ H-\mu_\eta {j}_z$ and $H-\mu_\eta \,\text{sign}({j}_z)$ and the
corresponding value in Fig.~\ref{fig:en} corresponding to the
constraint $H-\mu_\eta\sigma_z$.  For $N=20$ one obtains at $\mu_\eta=1$ the
values 64 and 17, and 96 and 25 for $\mu_\eta=2$ respectively for the type
of constraints discussed in Section~\ref{sec:nuclei}.  These values
should be compared to the maximum value $K_N(\mu_\eta) \approx 13$ and
lower values for the $N=20$ ``neutron'' system for large values of
$\eta$ in Fig.~\ref{fig:en}. Note that $K_N(\mu_\eta)$ is not a monotonous
function of $\mu_\eta$, see Fig.~\ref{fig:en}. One can now better appreciate
the difference between enforcing a constraint either on ${j}_z$ or
on $\text{sign}({j}_z)$, which imparts a relatively large angular
momentum to the system, while also by default leading to its
spin-polarization. The spin-polarization of the system obtained in this manner
is an effect caused by the finite angular momentum, per discussion in
Section~\ref{sec:nuclei}. For each $N$ the system reached it maximum
possible spin-polarization $\Sigma_\text{max} = |N_\uparrow-N_\downarrow| = N$.
The constraint on $\sigma_z$ is directly related to the generalized
density $n_{\sigma,\sigma'}({\bm r})$, which is naturally related to
$N_{\sigma}= \int d^3r\; n_{\sigma,\sigma}({\bm r})$ and thus to the
spin-polarization of the system $\Sigma=| N_\uparrow - N_\downarrow| $. In this case
the finite spin-polarization, namely the cause, induces a finite total angular
momentum, which is thus an effect.

While the many-body states thus obtained are characterized by axial symmetry, 
and thus by a good value of the 
projection of the total angular momentum on the symmetry axis, these states break 
rotational invariance, as these states do not have a well defined total angular momentum.  
But so do Slater determinants, formed with single-particle wave functions from an open shell  
One can denote these states $K\Sigma$-isomers and after 
angular momentum projection~\cite{Ring:2004} one can restore the rotational symmetry.
The spin asymmetry $\Sigma$ along the system symmetry axis is not affected by angular 
momentum projection, similarly to the $K$ quantum number, 
since $[j_z,\sigma_z]=0$ and $[\sum_{k=1}^Nj_z(k),\sum_{l=1}^N\sigma_z(l)]=0$. 
Notice, see Figs.~\ref{fig:en}, 
\ref{fig:sigma}, and \ref{fig:ES},
that even though $\Sigma$ does not necessarily acquire integer values always, it is in a 
matter of speaking ``quantized,'' as it changes by clearly defined jumps. 
It would be interesting to establish to what extent these kind of 
$\Sigma$-jumps survive upon the onset of pairing correlations. 
These $K\Sigma$-isomers  are by construction the lowest states 
characterized by a good quantum number $K$ and a defined spin-asymmetry.  

\section{Conclusions} 

I discussed here several frameworks suggested in the literature, as
well as a few new ones, designed to treat odd fermion superfluid
systems, their advantages and disadvantages, and I introduced a two
chemical potential
framework~\cite{Bulgac:2007,aslda,Sensarma:2007,Bulgac:2008}, which
appears to be much simpler to implement in comparison with other
frameworks suggested so far in the nuclear physics literature. 

Unlike the general framework designed by \textcite{Dobaczewski:1997},
which requires an {\it a priori} knowledge of the quantum numbers
characterizing the extra add fermion, the two chemical potentials
framework outlined in Section~\ref{sec:best} and when properly
numerically formulated, see Appendix~\ref{sec:numeric}, eschews the
diagonalization of the quasiparticle Hamiltonian, used in order to
arrive at a self-consistent solution, and also the need to identify
{\it a priori} the quantum numbers of the extra fermion. The
contribution of the lowest energy quasiparticle state of the odd
fermion is automatically selected in this framework, particularly when
the simulated annealing method is also incorporated in the iterative
process.  The gradient method advocated by \textcite{Robledo:2011}
apparently also does not require {\it a priori} identification of
the quantum numbers of the odd fermion, a method limited however only
to the ground states of odd or odd-odd nuclei. The two chemical
approach as described here can prove to be essential in constructing
the initial states in an induced fission process of a nucleus produced
in a many-nucleon transfer reaction, when the state of spin-polarization
can be arbitrary.  One can also generalize the approach in a
straightforward manner to impose both a finite spin-polarization and a
total angular momentum as well, to obtained what I have called above \
$K\Sigma$-isomers. Since the spin-polarization $\Sigma$ can be
``quantized,'' the identification of $K\Sigma$-isomers can lead to a deeper 
understanding of nuclear dynamics, e.g. in fission, and of the 
properties of nuclear spectra, in a manner analogous to the role
played by $K$-isomers.

At this time it is not clear whether either the simulated annealing
method or the gradient method~\cite{Robledo:2011} for generating the
new mean field during the iterative process is superior
computationally, or even if a carefully combination of the simulated
annealing with the gradient method could prove to be the best choice.
Since there exists methods~\cite{jin:2017,Kashiwaba:2020} which eschew
the ubiquitous diagonalization of the quasiparticle Hamiltonian used
routinely in the literature, these methods can and should be used in
conjunction with simulated annealing and/or gradient methods, in the
mean field calculations of polarized fermion systems and then one can
achieve a significant numerical speed-up.  The framework described
here is formally identical to describing nuclei in the presence of
strong magnetic fields, which one can encounter in magnetars for
example. In the presence of arbitrary strong magnetic fields one can
fully polarize a nucleus.

I have conjectured that by studying the asymmetry of the fission
fragments distributions, emitted along the direction of the angular
momentum of the fissioning odd or odd-odd nucleus one could shed new
light on whether time-odd mean field components play a qualitative new
role in fission. Compound states in fissioning nuclei with relatively
large angular momenta can be populated in one or many neutron transfer
reactions~\cite{Ramos:2020}.

Similar, but not identical, type of correlations have been recently
analyzed within the phenomenological model CGMF~\cite{Becker:2013}
based on the Hauser-Feshbach framework~\cite{Hauser:1952} by
\textcite{Lovell:2020} in neutron induced fission of actinides.  These
authors, while pointing to quite a number of experimental results,
observe that with the increasing energy of the incident neutron the
anisotropy for the reaction U$^{238}$(n,f) is noticeably more
pronounced than for the reactions U$^{235}$(n,f) and
Pu$^{239}$(n,f). This analysis thus suggests that fission is favored
along the direction of the total angular momentum of the compound
nucleus. This type of anisotropy has been postulated by
\textcite{Bohr:1956}.  It is not clear yet however, whether the
emission of the heavy fission fragment is favored or hindered over the
emission of the light fission fragment along the direction of the
total angular momentum of the compound nucleus.

\section {acknowledgement}

I thank G.F. Bertsch for a stimulating discussion and I. Stetcu for me
making me aware of the recent phenomenological study of fission
fragments anisotropy~\cite{Lovell:2020}.  I also thank J.E. Drut for
making a number of useful suggestions on the manuscript.  The work was
supported by U.S. Department of Energy, Office of Science, Grant
No. DE-FG02-97ER41014 and in part by NNSA cooperative agreement
DE-NA0003841.

\appendix
\section{Representation of the Pauli matrices }\label{sec:Pauli}

{There is a formal aspect that is hardly ever discussed in the
literature, the choice of the three axes for the spin and their
relation with the actual spatial directions. It is easy to verify that
the Schr\"odinger equation for a spin particle is invariant with
respect to the choice of the spin basis set, which is not the case in
the case of cold atom systems. Namely the Sch\"odinger equation does
not change if one replaces the canonical Pauli matrices with the new
Pauli matrices $\tilde {\bm \sigma}$
\begin{align}
\tilde{\bm \sigma}=  \exp\left ( - i\frac{\kappa}{2} {\bm \sigma}\cdot {\bm m} \right )  
{\bm \sigma}\exp\left ( i\frac{\kappa}{2} {\bm \sigma}\cdot {\bm m} \right ) , \label{eq:Pauli}
\end{align}  
where ${\bm m}=(\sin\chi\cos \zeta,\sin\chi\sin\zeta,\cos\chi)$ is
an arbitrary 3D unit vector and ${\bm \sigma}$ are the canonical Pauli
matrices.  The arbitrary angles $\kappa,\chi,\zeta$ have no relation
with the usual Euler angles $\psi,\theta, \phi$ describing a rotation
in the usual 3D space.  Basically what this means is that the
particular choice of $\tilde{\sigma}_z$ as a diagonal traceless Hermitian
matrix is not unique, and any traceless $2\times2$ Hermitian matrix
with eigenvalues $\pm 1$ is an equally acceptable choice. The same
applies for the other two Pauli matrices $\tilde{\sigma}_{x,y}$, with the only
requirement that 
\begin{align} 
[\tilde{\sigma}_k,\tilde{\sigma}_l]=2i\varepsilon_{klm}\tilde{\sigma}_m,
\end{align}
where $k,l,m=x,y,z$ and $\varepsilon_{klm}$ is the Levi-Civita symbol.
One example of a different choice for the Pauli spin matrices
different from the canonical one is
\begin{align}
&\tilde{\sigma}_x =  \begin{pmatrix} 0 & -i \\  i &   0 \end{pmatrix}, \;
&\tilde{\sigma}_y =  \begin{pmatrix} 1 & 0 \\ 0 & -1 \end{pmatrix}, \;
&\tilde{\sigma}_z =  \begin{pmatrix} 0 & 1 \\ 1 &  0 \end{pmatrix},
\end{align}
or any other even permutation of $\sigma_{x,y,z}$. One can choose also an odd permutation, 
but then the sign of one of the matrices has to be changed. 
Naturally, the single-particle angular momentum is then
\begin{align}
{\bm j}={\bm l}+\frac{\hbar}{2}\tilde{\bm \sigma}.
\end{align}

\section{Aspects of Numerical Implementation}\label{sec:numeric}

If the axial symmetry is codified explicitly into the numerical
single-particle basis used, then the action of the operator ${\cal
S}_z$ discussed in Section~\ref{sec:nuclei} is simply reduced to a
multiplication with $\eta =\text{sign}(m)=\pm 1$ and levels with
positive/negative $m$ quantum numbers are characterized by different
chemical potentials $\mu_m=\mu+\eta \mu_\eta$. If instead one uses the
operator $i\exp(-i\pi{j}_z/\hbar)$, then $\eta = i\exp(i\pi m)=\pm 1$ as
well, but the assignment of the quasiparticle states to either $N_\pm$
groups is different, as discussed in Section~\ref{sec:nuclei}.  If the
time-reversed orbitals with $+m$ and $-m$, which are typically
involved in the formation of Cooper pairs, when one of them is missing
in an odd system this leads to the spin-polarization of the nucleus, as
$n_+({\bm r},\sigma,\sigma')\ne n_-({\bm r},\sigma,\sigma')$.  \\
 
 Numerical implementation of the operators ${\cal S}_z,
i\exp(-i\pi{j}_x/\hbar)P, i\exp(-i\pi{j}_z/\hbar)$ on a 3D spatial lattice
however can run into numerical inaccuracies, as either the axial
symmetry or the rotations can be implemented only approximately.  In
this respect the use of operator ${\cal S}_z$ is preferable, as one
can use as an alternative method to determine $\eta$ the computation
of the sign of the expectation value of ${J}_z$, thus
\begin{align} \eta = \text{sign} \left (\langle
v_{km}|{J}_z|v_{km}\rangle \right ).
 \end{align} This method of introducing the $\mu_\eta{\cal S}_z$ into
the equations is equivalent to expressing the quasiparticle eigenvalue
as an expectation value of the quasiparticle Hamiltonian ${\cal
H}_{km}$ in the equations, see Eq.~\eqref{eq:eqp_static}. Namely one
can replace the equation ${\cal H}_{km}\psi_{km}=E_{km}\psi_{km}$ with
the mathematically equivalent equation ${\cal
H}_{km}\psi_{km}=\langle\psi_{km}|{\cal H}_{km}|\psi_{km}\rangle
\psi_{km}$.  The use of the framework discussed in
Section~\ref{sec:best} is clearly the simplest and unambiguous one.
 
 In the time-dependent problem there is no need to include the
operators $Q$, ${\cal S}_z$, or $\sigma_z$ and the simulation will
run as usual, even when symmetries  are broken during evolution.

The presence of an odd fermion on top of an even core leads to the
density spin-polarization of the even fermion core and also can lead to
time-reversal symmetry breaking in the ground state of such a system.
The determination of the eigenvalues and of the eigenvectors of
Eq.~\eqref{eq:nuclear} when many/all symmetries are broken and
particularly for the needs of time-dependent simulations, is an
extremely computationally demanding problem. This problem is
significantly simplified by noticing that the stationary solution is
determined fully by the densities alone, and the explicit presence of
quasiparticle wave functions is not required.  What this actually
means is that knowing the EDF, from which the quasiparticle
Hamiltonian ${\cal H}$ is extracted, various densities are simply
contour integrals of the resolvent $n_\sigma(r) = \oint_{\cal C} dz\;
1/(z-{\cal H})_{\sigma,\sigma} $, where the contour ${\cal C}$ is
appropriately chosen~\cite{jin:2017,Kashiwaba:2020}, and thus the
theory is practically orbital free, as in the original formulation of
the DFT by \textcite{Hohenberg:1964}, as opposed to the Kohn-Sham
formulation~\cite{Kohn:1965fk}.  The costly quasiparticle Hamiltonian
diagonalization can then be replaced by a significantly faster
algorithm, either the shifted conjugate orthogonal conjugate gradient
method~\cite{jin:2017} or the conjugate orthogonal conjugate residual
method~\cite{Kashiwaba:2020}. As during the iterative process the
lowest quasiparticle particle levels change their positions, sometimes
quite dramatically and their ordering can be significantly modified, a
simulated annealing method in conjunction with the iterative process
can lead to a significant stabilization of the iterations and the need
to specify in advance a specific quasiparticle state $\mu$ is then
eschewed. Within a simulated annealing process the iterative process
starts at a finite temperature, which is lowered according to a
predetermined schedule, and as the iterative process starts
converging, eventually the temperature reaches the goal $T=0$, where
$T$ is the temperature.  Figuratively, the black zero energy line in
Fig.~\ref{fig:Eqp} gets blurred and acquires a finite width $\sim{\cal
O}(T)$. In this manner quasiparticle states with both positive and
negative energies $E_k$ in a narrow energy band $|E_k|\sim{\cal O}(T)$
contribute to the densities and the somewhat erratic behavior of low
energy quasiparticle energies during the iterative process is
mitigated.  The simulated annealing method has been applied
successfully to cold atom
systems~\cite{Bulgac:2007,Bulgac:2008,Bulgac:2011a,aslda}.

The iterative process can likely be further accelerated also by using the 
gradient method advocated by \textcite{Robledo:2011}, 
which can be used in conjunction with the methods proposed by \textcite{jin:2017,Kashiwaba:2020}.   
After the iterative process converged one has to perform only a single diagonalization in order to determine the quasiparticle wave functions, 
which are needed as input for time-dependent  
simulations~\cite{Bulgac:2016,Bulgac:2017,Bulgac:2019a,Bulgac:2019b,Bulgac:2020,Wlazlowski:2016,Barton:2020}.

\providecommand{\selectlanguage}[1]{}
\renewcommand{\selectlanguage}[1]{}

\bibliography{latest_fission-IS}

\end{document}